\documentclass{article}%
\usepackage{amsmath}
\usepackage{amsfonts}
\usepackage{amssymb}
\usepackage{graphicx}
\usepackage[USenglish]{babel}%
\newtheorem{theorem}{Theorem}
\newtheorem{acknowledgement}[theorem]{Acknowledgement}

\begin{document}

\author{E. CAPELAS de OLIVEIRA and W. A. RODRIGUES Jr.\thanks{\textit{Int. J. Mod.
Phys. D} \textbf{13}(8), 1637-1659 (2004)}\\\hspace{-0.1cm}Institute of Mathematics, Statistics and Scientific Computation\\IMECC-UNICAMP CP 6065\\13083-970 Campinas-SP, Brazil\\walrod@ime.unicamp.br\hspace{0.15in}capelas@ime.unicamp.br}
\title{Dotted and Undotted Algebraic Spinor Fields in General Relativity}
\date{April, 17 2004\\
revised June, 17 2004\\
revised: August 05 2004\\
last revised: \ September 23 2004}
\maketitle

\begin{abstract}
We investigate using Clifford algebra methods the theory of \textit{algebraic}
dotted and undotted spinor fields over a Lorentzian spacetime and their
realizations as matrix spinor fields, which are the usual dotted and undotted
two component spinor fields. We found that some ad hoc rules postulated for
the covariant derivatives of Pauli sigma matrices and also for the Dirac gamma
matrices \ in General Relativity cover important physical meaning, which is
not apparent in the usual matrix presentation of the theory of two components
dotted and undotted spinor fields. We also discuss some issues related to the
previous one and which appear in a proposed "unified" \ theory of gravitation
and electromagnetism which use two components dotted and undotted spinor
fields and also \textit{paravector} fields, which are particular sections of
the even subundle of the Clifford bundle of spacetime.

\end{abstract}

\section{Introduction}

In this paper, using the general theory of Clifford and spin-Clifford bundles,
as described in \cite{moro,28} we scrutinize the concept of covariant
derivatives of \textit{algebraic} \textit{dotted} and \textit{undotted} spinor
fields,\footnote{These objects in our formalism, are represented as sections
of some well defined real spinor bundles, which are particular cases of a
general spin-Clifford bundle. We recall that the concept of real spinor fields
have been introduced by Hestenes in \cite{hestenesspinors}, but a rigorous
theory of that objects in a Lorentzian spacetime has only recently been
achieved \cite{moro,28}.} which have as matrix representatives the standard
two components spinor fields (dotted and undotted) already introduced long
ago, see, e.g., \cite{carmeli,penrose, penrindler, pirani}. What is
\textit{new} here is that we identify in the theory of algebraic spinor fields
an important and nontrivial physical interpretation for some
\textit{postulated} rules that are used in the standard formulation of the
\textit{matrix} spinor fields, e.g., why the covariant derivative of the Pauli
matrices must be null. We show that such a rule implies some constraints on
the geometry of the spacetime manifold, with admit a very interesting
geometrical interpretation. Indeed, \ a possible realization of that rules in
the Clifford bundle formalism is one where the vector fields defining a global
tetrad $\{\mathbf{e}_{\mathbf{a}}\}$ must be such that $D_{\mathbf{e}%
_{\mathbf{0}}}\mathbf{e}_{\mathbf{a}}=0$, i.e., $\mathbf{e}_{\mathbf{0}}$ must
be a geodesic reference frame and along each one of its integral lines, say
$\sigma,$ the $\mathbf{e}_{\mathbf{d}}$ $(\mathbf{d}=1,2,3)$ must be Fermi
transported, i.e., they are not rotating relative to the local gyroscope axes.
For the best of our knowledge these important facts are here disclosed for the
first time. We also examine the genesis of some ad hoc rules that are
postulated for the covariant derivatives of some \textit{paravector}
fields\footnote{In \cite{s1,s2,s3} the author states that the basic variables
of his \ `unified' theory \ are quaternion fields over a Lorentzian spacetime.
Well, they are \textit{not} are will be proved below.} \cite{s1,s2,s3} in some
proposed \ `unified' theories and for the Dirac gamma matrices in General
Relativity \cite{choquet}.

\section{Spacetime, Pauli and Quaternion Algebras}

In this section we recall some facts concerning three special \textit{real}
Clifford algebras, namely, the \textit{spacetime} algebra $\mathbb{R}_{1,3}$,
the \textit{Pauli} algebra $\mathbb{R}_{3,0}$ and the \textit{quaternion}
algebra $\mathbb{R}_{0,2}=\mathbb{H}$ and the relation between
them.\footnote{This material is treated in details e.g, in the books
\cite{cru,lounesto,porteous1,porteous2}. See also
\cite{femoro101,femoro201,femoro301,femoro401,femoro501,femoro601,femoro701}.}

\subsection{Spacetime Algebra}

To start, we recall that the spacetime algebra $\mathbb{R}_{1,3}$ is the
\textit{real }Clifford algebra associated \ with Minkowski vector space
$\mathbb{R}^{1,3}$, which is a four dimensional real vector space, equipped
with a Lorentzian bilinear form
\begin{equation}%
%TCIMACRO{\TeXButton{eta}{\mbox{\boldmath{$\eta$}}}}%
%BeginExpansion
\mbox{\boldmath{$\eta$}}%
%EndExpansion
:\mathbb{R}^{1,3}\times\mathbb{R}^{1,3}\rightarrow\mathbb{R}. \label{1}%
\end{equation}

Let $\{\mathbf{m}_{0}\mathbf{,m}_{1}\mathbf{,m}_{2}\mathbf{,m}_{3}\}$ be an
arbitrary orthonormal basis of $\mathbb{R}^{1,3}$, i.e.,%
\begin{equation}%
%TCIMACRO{\TeXButton{eta}{\mbox{\boldmath{$\eta$}}}}%
%BeginExpansion
\mbox{\boldmath{$\eta$}}%
%EndExpansion
(\mathbf{m}_{\mu}\mathbf{,m}_{\nu})=\eta_{\mu\nu}=\left\{
\begin{array}
[c]{ccc}%
1 & \text{if} & \mu=\nu=0\\
-1 & \text{if} & \mu=\nu=1,2,3\\
0 & \text{if} & \mu\neq\nu
\end{array}
\right.  \label{2}%
\end{equation}
As usual we resume Eq.(\ref{2}) writing $\mathbf{\eta}_{\mu\nu}=\mathrm{diag}%
(1,-1,-1,-1)$. We denote by $\{\mathbf{m}^{0}\mathbf{,m}^{1}\mathbf{,m}%
^{2}\mathbf{,m}^{3}\}$ the \textit{reciprocal} basis of $\{\mathbf{m}%
_{0}\mathbf{,m}_{1}\mathbf{,m}_{2}\mathbf{,m}_{3}\}$, i.e., $%
%TCIMACRO{\TeXButton{eta}{\mbox{\boldmath{$\eta$}}}}%
%BeginExpansion
\mbox{\boldmath{$\eta$}}%
%EndExpansion
\mathbf{(m}^{\mu}\mathbf{,m}_{\nu})=\delta_{\nu}^{\mu}$. We have in obvious
notation $\mathbf{\eta}(\mathbf{m}^{\mu}\mathbf{,m}^{\nu})=\eta^{\mu\nu}=$
$\mathrm{diag}(1,-1,-1,-1)$.

The spacetime algebra $\mathbb{R}_{1,3}$ is generate by the following
algebraic fundamental relation%
\begin{equation}
\mathbf{m}^{\mu}\mathbf{m}^{\nu}\mathbf{+m}^{\nu}\mathbf{m}^{\mu}=2\eta
^{\mu\nu}. \label{3}%
\end{equation}

\bigskip

We observe that in the above formula and in all the text the Clifford product
is denoted by \textit{juxtaposition} of symbols. The spacetime algebra
$\mathbb{R}_{1,3}$ as a vector space over the real field is isomorphic to the
exterior algebra $%
%TCIMACRO{\dbigwedge }%
%BeginExpansion
{\displaystyle\bigwedge}
%EndExpansion
\mathbb{R}^{1,3}=%
%TCIMACRO{\dsum \limits_{j=0}^{4}}%
%BeginExpansion
{\displaystyle\sum\limits_{j=0}^{4}}
%EndExpansion
$ $%
%TCIMACRO{\dbigwedge \nolimits^{j}}%
%BeginExpansion
{\displaystyle\bigwedge\nolimits^{j}}
%EndExpansion
\mathbb{R}^{1,3}$ of $\mathbb{R}^{1,3}$. We code that information writing $%
%TCIMACRO{\dbigwedge }%
%BeginExpansion
{\displaystyle\bigwedge}
%EndExpansion
\mathbb{R}^{1,3}\hookrightarrow\mathbb{R}_{1,3}$. \ Also, we make the
following identifications: $%
%TCIMACRO{\dbigwedge \nolimits^{0}}%
%BeginExpansion
{\displaystyle\bigwedge\nolimits^{0}}
%EndExpansion
\mathbb{R}^{1,3}\equiv\mathbb{R}$ and $%
%TCIMACRO{\dbigwedge \nolimits^{1}}%
%BeginExpansion
{\displaystyle\bigwedge\nolimits^{1}}
%EndExpansion
\mathbb{R}^{1,3}\equiv\mathbb{R}^{1,3}$ . Moreover, we identify the exterior
product of vectors by
\begin{equation}
\mathbf{m}^{\mu}\mathbf{\wedge m}^{\nu}\mathbf{=}\frac{1}{2}\left(
\mathbf{m}^{\mu}\mathbf{m}^{\nu}\mathbf{-m}^{\nu}\mathbf{m}^{\mu}\right)  ,
\label{4}%
\end{equation}
and also, we identify the scalar product of of vectors by
\begin{equation}%
%TCIMACRO{\TeXButton{eta}{\mbox{\boldmath{$\eta$}}}}%
%BeginExpansion
\mbox{\boldmath{$\eta$}}%
%EndExpansion
\mathbf{(m}^{\mu}\mathbf{,m}^{\nu}\mathbf{)=}\frac{1}{2}\left(  \mathbf{m}%
^{\mu}\mathbf{m}^{\nu}\mathbf{+m}^{\nu}\mathbf{m}^{\mu}\right)  . \label{5}%
\end{equation}
Then we can write
\begin{equation}
\mathbf{m}^{\mu}\mathbf{m}^{\nu}=%
%TCIMACRO{\TeXButton{eta}{\mbox{\boldmath{$\eta$}}}}%
%BeginExpansion
\mbox{\boldmath{$\eta$}}%
%EndExpansion
\mathbf{(m}^{\mu}\mathbf{,m}^{\nu}\mathbf{)+m}^{\mu}\mathbf{\wedge m}^{\nu}.
\label{6}%
\end{equation}
\ Now, an arbitrary element $\mathbf{C}\in\mathbb{R}_{1,3}$ can be written as
sum of \textit{nonhomogeneous multivectors}, i.e.,%
\begin{equation}
\mathbf{C}=s+c_{\mu}\mathbf{m}^{\mu}\mathbf{+}\frac{1}{2}c_{\mu\nu}%
\mathbf{m}^{\mu}\mathbf{m}^{\nu}+\frac{1}{3!}c_{\mu\nu\rho}\mathbf{m}^{\mu
}\mathbf{m}^{\nu}\mathbf{m}^{\rho}+p\mathbf{m}^{5} \label{7}%
\end{equation}
where $s,c_{\mu},c_{\mu\nu},c_{\mu\nu\rho},p\in\mathbb{R}$ and $c_{\mu\nu
},c_{\mu\nu\rho}$ are completely antisymmetric in all indices. Also
$\mathbf{m}^{5}\mathbf{=m}^{0}\mathbf{m}^{1}\mathbf{m}^{2}\mathbf{m}^{3}$ is
the generator of the pseudo scalars. As matrix algebra we have that
$\mathbb{R}_{1,3}\simeq\mathbb{H(}2)$, the algebra of the $2\times2$
quaternionic matrices.

\subsection{Pauli Algebra}

Now, we recall that the Pauli algebra $\mathbb{R}_{3,0}$ is the real Clifford
algebra associated with the Euclidean vector space $\mathbb{R}^{3,0}$,
equipped as usual, with a positive definite bilinear form. \ As a matrix
algebra we have that $\mathbb{R}_{3,0}\simeq\mathbb{C}\left(  2\right)  $, the
algebra of $2\times2$ complex matrices. Moreover, we recall that
$\mathbb{R}_{3,0}$ is isomorphic to the even subalgebra of the spacetime
algebra, i.e., writing $\mathbb{R}_{1,3}=$ $\mathbb{R}_{1,3}^{(0)}\oplus$
$\mathbb{R}_{1,3}^{(1)}$ we have,
\begin{equation}
\mathbb{R}_{3,0}\simeq\mathbb{R}_{1,3}^{(0)}. \label{8}%
\end{equation}

The isomorphism is easily exhibited by putting $%
%TCIMACRO{\TeXButton{sigma}{\mbox{\boldmath{$\sigma$}}}}%
%BeginExpansion
\mbox{\boldmath{$\sigma$}}%
%EndExpansion
^{i}\mathbf{=m}^{i}\mathbf{m}^{0}$, $i=1,2,3$. Indeed, $\ $with $\delta
^{ij}=\mathrm{diag}(1,1,1)$, we have
\begin{equation}%
%TCIMACRO{\TeXButton{sigma}{\mbox{\boldmath{$\sigma$}}}}%
%BeginExpansion
\mbox{\boldmath{$\sigma$}}%
%EndExpansion
^{i}%
%TCIMACRO{\TeXButton{sigma}{\mbox{\boldmath{$\sigma$}}}}%
%BeginExpansion
\mbox{\boldmath{$\sigma$}}%
%EndExpansion
^{j}\mathbf{+}%
%TCIMACRO{\TeXButton{sigma}{\mbox{\boldmath{$\sigma$}}}}%
%BeginExpansion
\mbox{\boldmath{$\sigma$}}%
%EndExpansion
^{j}%
%TCIMACRO{\TeXButton{sigma}{\mbox{\boldmath{$\sigma$}}}}%
%BeginExpansion
\mbox{\boldmath{$\sigma$}}%
%EndExpansion
^{i}=2\delta^{ij}, \label{9}%
\end{equation}
which is the fundamental relation defining the algebra $\mathbb{R}_{3,0}$.
Elements of the Pauli algebra will be called Pauli numbers\footnote{Sometimes
they are also called `complex quaternions'. This last terminology will be
obvious in a while.}. As vector space over the real field, we have that
$\mathbb{R}_{3,0}\ $is isomorphic to$%
%TCIMACRO{\dbigwedge }%
%BeginExpansion
{\displaystyle\bigwedge}
%EndExpansion
\mathbb{R}^{3,0}\hookrightarrow\mathbb{R}_{3,0}\subset\mathbb{R}_{1,3}$. So,
any Pauli number can be written as
\begin{equation}
\mathbf{P}=s+p^{i}%
%TCIMACRO{\TeXButton{sigma}{\mbox{\boldmath{$\sigma$}}}}%
%BeginExpansion
\mbox{\boldmath{$\sigma$}}%
%EndExpansion
^{i}+\frac{1}{2}p_{ij}^{i}%
%TCIMACRO{\TeXButton{sigma}{\mbox{\boldmath{$\sigma$}}}}%
%BeginExpansion
\mbox{\boldmath{$\sigma$}}%
%EndExpansion
^{i}%
%TCIMACRO{\TeXButton{sigma}{\mbox{\boldmath{$\sigma$}}}}%
%BeginExpansion
\mbox{\boldmath{$\sigma$}}%
%EndExpansion
^{j}+p\text{\textsc{i}}\mathbf{,} \label{10}%
\end{equation}
where $s,p_{i},p_{ij},p\in\mathbb{R}$ and $p_{ij}=-p_{ji}$ and also
\begin{equation}
\text{\textsc{i}}\mathbf{=}%
%TCIMACRO{\TeXButton{sigma}{\mbox{\boldmath{$\sigma$}}}}%
%BeginExpansion
\mbox{\boldmath{$\sigma$}}%
%EndExpansion
^{1}%
%TCIMACRO{\TeXButton{sigma}{\mbox{\boldmath{$\sigma$}}}}%
%BeginExpansion
\mbox{\boldmath{$\sigma$}}%
%EndExpansion
^{2}%
%TCIMACRO{\TeXButton{sigma}{\mbox{\boldmath{$\sigma$}}}}%
%BeginExpansion
\mbox{\boldmath{$\sigma$}}%
%EndExpansion
^{3}=\mathbf{m}^{5}. \label{11}%
\end{equation}

\bigskip\bigskip Note that \textsc{i}$^{2}=-1$ and that \textsc{i} commutes
with any Pauli number. We can trivially verify%
\begin{align}%
%TCIMACRO{\TeXButton{sigma}{\mbox{\boldmath{$\sigma$}}}}%
%BeginExpansion
\mbox{\boldmath{$\sigma$}}%
%EndExpansion
^{i}%
%TCIMACRO{\TeXButton{sigma}{\mbox{\boldmath{$\sigma$}}}}%
%BeginExpansion
\mbox{\boldmath{$\sigma$}}%
%EndExpansion
^{j}  &  =\text{\textsc{i}}\varepsilon_{k}^{i\text{ }j}%
%TCIMACRO{\TeXButton{sigma}{\mbox{\boldmath{$\sigma$}}}}%
%BeginExpansion
\mbox{\boldmath{$\sigma$}}%
%EndExpansion
^{k}+\delta^{ij},\label{12}\\
\mathbf{[}%
%TCIMACRO{\TeXButton{sigma}{\mbox{\boldmath{$\sigma$}}}}%
%BeginExpansion
\mbox{\boldmath{$\sigma$}}%
%EndExpansion
^{i}\mathbf{,}%
%TCIMACRO{\TeXButton{sigma}{\mbox{\boldmath{$\sigma$}}}}%
%BeginExpansion
\mbox{\boldmath{$\sigma$}}%
%EndExpansion
^{j}\mathbf{]}  &  \mathbf{\equiv}%
%TCIMACRO{\TeXButton{sigma}{\mbox{\boldmath{$\sigma$}}}}%
%BeginExpansion
\mbox{\boldmath{$\sigma$}}%
%EndExpansion
^{i}%
%TCIMACRO{\TeXButton{sigma}{\mbox{\boldmath{$\sigma$}}}}%
%BeginExpansion
\mbox{\boldmath{$\sigma$}}%
%EndExpansion
^{j}\mathbf{-}%
%TCIMACRO{\TeXButton{sigma}{\mbox{\boldmath{$\sigma$}}}}%
%BeginExpansion
\mbox{\boldmath{$\sigma$}}%
%EndExpansion
^{j}%
%TCIMACRO{\TeXButton{sigma}{\mbox{\boldmath{$\sigma$}}}}%
%BeginExpansion
\mbox{\boldmath{$\sigma$}}%
%EndExpansion
^{i}\mathbf{=}2%
%TCIMACRO{\TeXButton{sigma}{\mbox{\boldmath{$\sigma$}}}}%
%BeginExpansion
\mbox{\boldmath{$\sigma$}}%
%EndExpansion
^{i}\mathbf{\wedge}%
%TCIMACRO{\TeXButton{sigma}{\mbox{\boldmath{$\sigma$}}}}%
%BeginExpansion
\mbox{\boldmath{$\sigma$}}%
%EndExpansion
^{j}=2\text{\textsc{i}}\varepsilon_{k}^{i\text{ }j}%
%TCIMACRO{\TeXButton{sigma}{\mbox{\boldmath{$\sigma$}}}}%
%BeginExpansion
\mbox{\boldmath{$\sigma$}}%
%EndExpansion
^{k}.\nonumber
\end{align}

\bigskip

In that way, writing $\mathbb{R}_{3,0}=\mathbb{R}_{3,0}^{(0)}+\mathbb{R}%
_{3,0}^{(1)}$, any Pauli number can be written as%
\begin{equation}
\mathbf{P=Q}_{1}\mathbf{+}\text{\textsc{i}}\mathbf{Q}_{2},\hspace
{0.15in}\mathbf{Q}_{1}\in\mathbb{R}_{3,0}^{(0)},\hspace{0.15in}%
\text{\textsc{i}}\mathbf{Q}_{2}\in\mathbb{R}_{3,0}^{(1)}, \label{13}%
\end{equation}
with
\begin{align}
\mathbf{Q}_{1}  &  =a_{0}+a_{k}(\text{\textsc{i}}%
%TCIMACRO{\TeXButton{sigma}{\mbox{\boldmath{$\sigma$}}}}%
%BeginExpansion
\mbox{\boldmath{$\sigma$}}%
%EndExpansion
^{k}),\hspace{0.15in}a_{0}=s,\hspace{0.15in}a_{k}=\frac{1}{2}\varepsilon
_{k}^{i\text{ }j}p_{ij},\label{14}\\
\mathbf{Q}_{2}  &  =\text{\textsc{i}}\left(  b_{0}+b_{k}(\text{\textsc{i}}%
%TCIMACRO{\TeXButton{sigma}{\mbox{\boldmath{$\sigma$}}}}%
%BeginExpansion
\mbox{\boldmath{$\sigma$}}%
%EndExpansion
^{k}\right)  ),\hspace{0.15in}b_{0}=p,\hspace{0.15in}b_{k}=-p_{k}.\nonumber
\end{align}

\subsection{Quaternion Algebra}

Eqs.(\ref{14}) \ show that the quaternion algebra $\mathbb{R}_{0,2}%
=\mathbb{H}$ can be identified as the even subalgebra of $\mathbb{R}_{3,0}$, i.e.,%

\begin{equation}
\mathbb{R}_{0,2}=\mathbb{H\simeq R}_{3,0}^{(0)}. \label{15}%
\end{equation}

The statement is obvious once we identify the basis $\{1,\mathit{\hat{\imath}%
},\mathit{\hat{\jmath}}\mathbf{,\mathit{\hat{k}}\}}$ of $\mathbb{H}$ with
\begin{equation}
\{\mathbf{1,}\text{\textsc{i}}%
%TCIMACRO{\TeXButton{sigma}{\mbox{\boldmath{$\sigma$}}}}%
%BeginExpansion
\mbox{\boldmath{$\sigma$}}%
%EndExpansion
^{1}\mathbf{,}\text{\textsc{i}}%
%TCIMACRO{\TeXButton{sigma}{\mbox{\boldmath{$\sigma$}}}}%
%BeginExpansion
\mbox{\boldmath{$\sigma$}}%
%EndExpansion
^{2}\mathbf{,}\text{\textsc{i}}%
%TCIMACRO{\TeXButton{sigma}{\mbox{\boldmath{$\sigma$}}}}%
%BeginExpansion
\mbox{\boldmath{$\sigma$}}%
%EndExpansion
^{3}\}, \label{15'}%
\end{equation}
which are the generators of $\mathbb{R}_{3,0}^{(0)}$. We observe moreover that
the even subalgebra of the quaternions can be identified (in an obvious way)
with the complex field, i.e., $\mathbb{R}_{0,2}^{(0)}\simeq\mathbb{C}$.

\ \ Returning to Eq.(\ref{10}) we see that any $\mathbf{P}\in\mathbb{R}_{3,0}$
can also be written as
\begin{equation}
\mathbf{P=P}_{1}\mathbf{+\text{\textsc{i}}L}_{2}, \label{16}%
\end{equation}
where
\begin{align}
\mathbf{P}_{1}  &  =(s+p_{k}%
%TCIMACRO{\TeXButton{sigma}{\mbox{\boldmath{$\sigma$}}}}%
%BeginExpansion
\mbox{\boldmath{$\sigma$}}%
%EndExpansion
^{k})\in%
%TCIMACRO{\dbigwedge \nolimits^{0}}%
%BeginExpansion
{\displaystyle\bigwedge\nolimits^{0}}
%EndExpansion
\mathbb{R}^{3,0}\oplus%
%TCIMACRO{\dbigwedge \nolimits^{1}}%
%BeginExpansion
{\displaystyle\bigwedge\nolimits^{1}}
%EndExpansion
\mathbb{R}^{3,0}\equiv\mathbb{R\oplus}%
%TCIMACRO{\dbigwedge \nolimits^{1}}%
%BeginExpansion
{\displaystyle\bigwedge\nolimits^{1}}
%EndExpansion
\mathbb{R}^{3,0},\nonumber\\
\text{\textsc{i}}\mathbf{L}_{2}  &  =\text{\textsc{i}}\mathbf{(}%
p+\text{\textsc{i}}l_{k}%
%TCIMACRO{\TeXButton{sigma}{\mbox{\boldmath{$\sigma$}}}}%
%BeginExpansion
\mbox{\boldmath{$\sigma$}}%
%EndExpansion
^{k})\in%
%TCIMACRO{\dbigwedge \nolimits^{2}}%
%BeginExpansion
{\displaystyle\bigwedge\nolimits^{2}}
%EndExpansion
\mathbb{R}^{3,0}\oplus%
%TCIMACRO{\dbigwedge \nolimits^{3}}%
%BeginExpansion
{\displaystyle\bigwedge\nolimits^{3}}
%EndExpansion
\mathbb{R}^{3,0}, \label{17}%
\end{align}
with $l_{k}=-\varepsilon_{k}^{i\text{ }j}p_{ij}\in\mathbb{R}$. The important
fact that we want to emphasize here is that the subspaces $(\mathbb{R\oplus}%
%TCIMACRO{\dbigwedge \nolimits^{1}}%
%BeginExpansion
{\displaystyle\bigwedge\nolimits^{1}}
%EndExpansion
\mathbb{R}^{3,0})$ and $(%
%TCIMACRO{\dbigwedge \nolimits^{2}}%
%BeginExpansion
{\displaystyle\bigwedge\nolimits^{2}}
%EndExpansion
\mathbb{R}^{3,0}\oplus%
%TCIMACRO{\dbigwedge \nolimits^{3}}%
%BeginExpansion
{\displaystyle\bigwedge\nolimits^{3}}
%EndExpansion
\mathbb{R}^{3,0})$ do not close separately any algebra. In general, if
$\mathbf{A,C}\in(\mathbb{R\oplus}%
%TCIMACRO{\dbigwedge \nolimits^{1}}%
%BeginExpansion
{\displaystyle\bigwedge\nolimits^{1}}
%EndExpansion
\mathbb{R}^{3,0})$ then%
\begin{equation}
\mathbf{AC}\in\mathbb{R\oplus}%
%TCIMACRO{\dbigwedge \nolimits^{1}}%
%BeginExpansion
{\displaystyle\bigwedge\nolimits^{1}}
%EndExpansion
\mathbb{R}^{3,0}\oplus%
%TCIMACRO{\dbigwedge \nolimits^{2}}%
%BeginExpansion
{\displaystyle\bigwedge\nolimits^{2}}
%EndExpansion
\mathbb{R}^{3,0}. \label{18}%
\end{equation}

\bigskip

To continue, we introduce
\begin{equation}%
%TCIMACRO{\TeXButton{sigma}{\mbox{\boldmath{$\sigma$}}}}%
%BeginExpansion
\mbox{\boldmath{$\sigma$}}%
%EndExpansion
_{i}\mathbf{=m}_{i}\mathbf{m}_{0}=\mathbf{-}%
%TCIMACRO{\TeXButton{sigma}{\mbox{\boldmath{$\sigma$}}}}%
%BeginExpansion
\mbox{\boldmath{$\sigma$}}%
%EndExpansion
^{i},\hspace{0.15in}i=1,2,3. \label{19}%
\end{equation}

\bigskip Then, \textsc{i}$=-%
%TCIMACRO{\TeXButton{sigma}{\mbox{\boldmath{$\sigma$}}}}%
%BeginExpansion
\mbox{\boldmath{$\sigma$}}%
%EndExpansion
_{1}%
%TCIMACRO{\TeXButton{sigma}{\mbox{\boldmath{$\sigma$}}}}%
%BeginExpansion
\mbox{\boldmath{$\sigma$}}%
%EndExpansion
_{2}%
%TCIMACRO{\TeXButton{sigma}{\mbox{\boldmath{$\sigma$}}}}%
%BeginExpansion
\mbox{\boldmath{$\sigma$}}%
%EndExpansion
_{3}$ and the basis $\{1,\hat{\imath}$\textit{ }$,\hat{\jmath},\hat
{k}\mathbf{\}}$ of $\mathbb{H}$ can be identified with $\{1,\mathbf{-}%
$\textsc{i}$%
%TCIMACRO{\TeXButton{sigma}{\mbox{\boldmath{$\sigma$}}}}%
%BeginExpansion
\mbox{\boldmath{$\sigma$}}%
%EndExpansion
_{1}\mathbf{,-}$\textsc{i}$%
%TCIMACRO{\TeXButton{sigma}{\mbox{\boldmath{$\sigma$}}}}%
%BeginExpansion
\mbox{\boldmath{$\sigma$}}%
%EndExpansion
_{2}\mathbf{,-}$\textsc{i}$%
%TCIMACRO{\TeXButton{sigma}{\mbox{\boldmath{$\sigma$}}}}%
%BeginExpansion
\mbox{\boldmath{$\sigma$}}%
%EndExpansion
_{3}\}$.

Now, we already said that $\mathbb{R}_{3,0}\simeq\mathbb{C}\left(  2\right)
$. This permit us to represent the Pauli numbers by \ $2\times2$ complex
matrices, in the usual way ($\mathrm{i}=\sqrt{-1}$). We write $\mathbb{R}%
_{3,0}\ni\mathbf{P}\mapsto P\in\mathbb{C(}2)$, with
\begin{equation}%
\begin{array}
[c]{ccc}%
%TCIMACRO{\TeXButton{sigma}{\mbox{\boldmath{$\sigma$}}}}%
%BeginExpansion
\mbox{\boldmath{$\sigma$}}%
%EndExpansion
^{1} & \mapsto & \sigma^{1}=\left(
\begin{array}
[c]{cc}%
0 & 1\\
1 & 0
\end{array}
\right) \\%
%TCIMACRO{\TeXButton{sigma}{\mbox{\boldmath{$\sigma$}}}}%
%BeginExpansion
\mbox{\boldmath{$\sigma$}}%
%EndExpansion
^{2} & \mapsto & \sigma^{2}=\left(
\begin{array}
[c]{cc}%
0 & -\mathrm{i}\\
\mathrm{i} & 0
\end{array}
\right) \\%
%TCIMACRO{\TeXButton{sigma}{\mbox{\boldmath{$\sigma$}}}}%
%BeginExpansion
\mbox{\boldmath{$\sigma$}}%
%EndExpansion
^{3} & \mapsto & \sigma^{3}=\left(
\begin{array}
[c]{cc}%
1 & 0\\
0 & -1
\end{array}
\right)  .
\end{array}
\label{20}%
\end{equation}

\subsection{Minimal left and right ideals in the Pauli Algebra and Spinors}

It is not our intention to present here the details of the general theory of
algebraic spinors. Nevertheless, we shall need \ to recall some results that
we necessary for what follows\footnote{For details, see, e.g.,
\cite{ficaro,moro,28}.}. The elements $\mathbf{e}_{\pm}=\frac{1}{2}(1+%
%TCIMACRO{\TeXButton{sigma}{\mbox{\boldmath{$\sigma$}}}}%
%BeginExpansion
\mbox{\boldmath{$\sigma$}}%
%EndExpansion
_{3})=\frac{1}{2}(1+\mathbf{m}_{3}\mathbf{m}_{0})\in\mathbb{R}_{1,3}%
^{(0)}\simeq\mathbb{R}_{3,0}$, $\mathbf{e}_{\pm}^{2}=\mathbf{e}_{\pm}$ are
\textit{minimal idempotents} of $\mathbb{R}_{3,0}$. They generate the minimal
left and right ideals%
\begin{equation}
\mathbf{I}_{\pm}=\mathbb{R}_{1,3}^{(0)}\mathbf{e}_{\pm},\hspace{0.15in}%
\mathbf{R}_{\pm}\mathbf{=e}_{\pm}\mathbb{R}_{1,3}^{(0)}. \label{25}%
\end{equation}

>From now on we write $\mathbf{e=e}_{+}$. It can be\ easily shown (see below)
that, e.g., $\mathbf{I=I}_{+}$ has the structure of a $2$-dimensional vector
space over the complex field \cite{ficaro,lounesto}, i.e., $\mathbf{I\simeq
}\mathbb{C}^{2}$. The elements of the vector space $\mathbf{I}$ are called
algebraic \textit{contravariant undotted spinors} and the elements of
$\mathbb{C}^{2}$ are the usual \ \textit{contravariant undotted spinors} used
in physics textbooks. They carry the $D^{(\frac{1}{2},0)\text{ }}$
representation of $Sl(2,\mathbb{C)}$ \cite{miller}. \ If \ $\mathbf{\varphi\in
I}$ we denote by $\varphi\in\mathbb{C}^{2}$ the usual matrix
representative\footnote{The matrix representation of the elements of the
ideals $\mathbf{I,\dot{I}}$, are of course, $2\times2$ complex matrices (see,
\cite{ficaro}, for details). It happens that both columns of that matrices
have the \textit{same} information and the representation by column matrices
is enough here for our purposes.} of $\mathbf{\varphi}$ is
\begin{equation}
\varphi=\left(
\begin{array}
[c]{c}%
\varphi^{1}\\
\varphi^{2}%
\end{array}
\right)  ,\hspace{0.15in}\varphi^{1},\varphi^{2}\in\mathbb{C}. \label{26}%
\end{equation}
\ We denote by $\mathbf{\dot{I}=e}\mathbb{R}_{1,3}^{(0)}$ the space of the
\ algebraic covariant dotted spinors. We have the isomorphism, $\mathbf{\dot
{I}\simeq(}\mathbb{C}^{2})^{\dagger}\simeq\mathbb{C}_{2}$, where $\dagger$
denotes Hermitian conjugation. The elements of $\mathbf{(}\mathbb{C}%
^{2})^{\dagger}$ are the usual contravariant spinor fields used in physics
textbooks. They carry the $D^{(0,\frac{1}{2})\text{ }}$ representation of
$Sl(2,\mathbb{C)}$ \cite{miller}. If \ $\overset{\cdot}{%
%TCIMACRO{\TeXButton{xi}{\mbox{\boldmath{$\xi$}}}}%
%BeginExpansion
\mbox{\boldmath{$\xi$}}%
%EndExpansion
}\in\mathbf{\dot{I}}$ its matrix representation \ in $\mathbf{(}\mathbb{C}%
^{2})^{\dagger}$ \ is a row matrix usually denoted by%
\begin{equation}
\dot{\xi}=\left(
\begin{array}
[c]{cc}%
\xi_{\dot{1}} & \xi_{\dot{2}}%
\end{array}
\right)  ,\hspace{0.15in}\xi_{\dot{1}},\xi_{\dot{2}}\in\mathbb{C}. \label{27}%
\end{equation}
The following representation of $\overset{\cdot}{%
%TCIMACRO{\TeXButton{xi}{\mbox{\boldmath{$\xi$}}}}%
%BeginExpansion
\mbox{\boldmath{$\xi$}}%
%EndExpansion
}\in\mathbf{\dot{I}}$ \ in $\mathbf{(}\mathbb{C}^{2})^{\dagger}$ is extremely
convenient. We say that to a covariant undotted spinor $\xi$ there corresponds
a covariant dotted spinor $\dot{\xi}$ given by
\begin{equation}
\mathbf{\dot{I}}\ni\overset{\cdot}{%
%TCIMACRO{\TeXButton{xi}{\mbox{\boldmath{$\xi$}}}}%
%BeginExpansion
\mbox{\boldmath{$\xi$}}%
%EndExpansion
}\mapsto\dot{\xi}=\bar{\xi}\varepsilon\in\mathbf{(}\mathbb{C}^{2})^{\dagger
},\hspace{0.15in}\bar{\xi}_{1},\bar{\xi}_{2}\in\mathbb{C}, \label{27'}%
\end{equation}
with%
\begin{equation}
\varepsilon=\left(
\begin{array}
[c]{cc}%
0 & 1\\
-1 & 0
\end{array}
\right)  . \label{27''}%
\end{equation}

\bigskip We can easily find a basis for $\mathbf{I}$ and $\mathbf{\dot{I}}$.
Indeed, since $\mathbf{I}=\mathbb{R}_{1,3}^{(0)}\mathbf{e}$ we have that any $%
%TCIMACRO{\TeXButton{bvarphi}{\mbox{\boldmath{$\varphi$}}}}%
%BeginExpansion
\mbox{\boldmath{$\varphi$}}%
%EndExpansion
\mathbf{\in I}$ can be written as%
\[%
%TCIMACRO{\TeXButton{bvarphi}{\mbox{\boldmath{$\varphi$}}}}%
%BeginExpansion
\mbox{\boldmath{$\varphi$}}%
%EndExpansion
\mathbf{=}%
%TCIMACRO{\TeXButton{bvarphi}{\mbox{\boldmath{$\varphi$}}}}%
%BeginExpansion
\mbox{\boldmath{$\varphi$}}%
%EndExpansion
^{1}%
%TCIMACRO{\TeXButton{vartheta}{\mbox{\boldmath{$\vartheta$}}}}%
%BeginExpansion
\mbox{\boldmath{$\vartheta$}}%
%EndExpansion
_{1}\mathbf{+}%
%TCIMACRO{\TeXButton{bvarphi}{\mbox{\boldmath{$\varphi$}}}}%
%BeginExpansion
\mbox{\boldmath{$\varphi$}}%
%EndExpansion
^{2}%
%TCIMACRO{\TeXButton{vartheta}{\mbox{\boldmath{$\vartheta$}}}}%
%BeginExpansion
\mbox{\boldmath{$\vartheta$}}%
%EndExpansion
_{2}%
\]
where
\begin{align}%
%TCIMACRO{\TeXButton{vartheta}{\mbox{\boldmath{$\vartheta$}}}}%
%BeginExpansion
\mbox{\boldmath{$\vartheta$}}%
%EndExpansion
_{1}  &  \mathbf{=}\mathbf{e,\hspace{0.15in}}%
%TCIMACRO{\TeXButton{vartheta}{\mbox{\boldmath{$\vartheta$}}}}%
%BeginExpansion
\mbox{\boldmath{$\vartheta$}}%
%EndExpansion
_{2}=%
%TCIMACRO{\TeXButton{sigma}{\mbox{\boldmath{$\sigma$}}}}%
%BeginExpansion
\mbox{\boldmath{$\sigma$}}%
%EndExpansion
_{1}\mathbf{e}\nonumber\\%
%TCIMACRO{\TeXButton{bvarphi}{\mbox{\boldmath{$\varphi$}}}}%
%BeginExpansion
\mbox{\boldmath{$\varphi$}}%
%EndExpansion
^{1}  &  =a+\mathbf{i}b,\hspace{0.15in}%
%TCIMACRO{\TeXButton{bvarphi}{\mbox{\boldmath{$\varphi$}}}}%
%BeginExpansion
\mbox{\boldmath{$\varphi$}}%
%EndExpansion
^{2}=c+\mathbf{i}d,\hspace{0.15in}a,b,c,d\in\mathbb{R.} \label{27A}%
\end{align}

Analogously we find that any $\overset{\cdot}{%
%TCIMACRO{\TeXButton{xi}{\mbox{\boldmath{$\xi$}}}}%
%BeginExpansion
\mbox{\boldmath{$\xi$}}%
%EndExpansion
}\in\mathbf{\dot{I}}$ can be written as
\begin{align}
\overset{\cdot}{%
%TCIMACRO{\TeXButton{xi}{\mbox{\boldmath{$\xi$}}}}%
%BeginExpansion
\mbox{\boldmath{$\xi$}}%
%EndExpansion
}  &  =%
%TCIMACRO{\TeXButton{xi}{\mbox{\boldmath{$\xi$}}}}%
%BeginExpansion
\mbox{\boldmath{$\xi$}}%
%EndExpansion
^{\dot{1}}\mathbf{s}^{\dot{1}}+%
%TCIMACRO{\TeXButton{xi}{\mbox{\boldmath{$\xi$}}}}%
%BeginExpansion
\mbox{\boldmath{$\xi$}}%
%EndExpansion
_{\dot{2}}\mathbf{s}^{\dot{2}}\nonumber\\
\mathbf{s}^{\dot{1}}  &  =\mathbf{e,}\hspace{0.15in}\mathbf{s}^{\dot{2}%
}=\mathbf{e}%
%TCIMACRO{\TeXButton{sigma}{\mbox{\boldmath{$\sigma$}}}}%
%BeginExpansion
\mbox{\boldmath{$\sigma$}}%
%EndExpansion
_{1}\mathbf{.} \label{27b}%
\end{align}

Defining the mapping
\begin{align}%
%TCIMACRO{\TeXButton{iota}{\mbox{\boldmath{$\iota$}}} }%
%BeginExpansion
\mbox{\boldmath{$\iota$}}
%EndExpansion
&  :\mathbf{I}\otimes\mathbf{\dot{I}\rightarrow}\mathbb{R}_{1,3}^{(0)}%
\simeq\mathbb{R}_{3,0},\nonumber\\%
%TCIMACRO{\TeXButton{iota}{\mbox{\boldmath{$\iota$}}}}%
%BeginExpansion
\mbox{\boldmath{$\iota$}}%
%EndExpansion
(%
%TCIMACRO{\TeXButton{varphi}{\mbox{\boldmath{$\varphi$}}}}%
%BeginExpansion
\mbox{\boldmath{$\varphi$}}%
%EndExpansion
\mathbf{\otimes}\overset{\cdot}{%
%TCIMACRO{\TeXButton{xi}{\mbox{\boldmath{$\xi$}}}}%
%BeginExpansion
\mbox{\boldmath{$\xi$}}%
%EndExpansion
})  &  =%
%TCIMACRO{\TeXButton{varphi}{\mbox{\boldmath{$\varphi$}}}}%
%BeginExpansion
\mbox{\boldmath{$\varphi$}}%
%EndExpansion
\overset{\cdot}{%
%TCIMACRO{\TeXButton{xi}{\mbox{\boldmath{$\xi$}}}}%
%BeginExpansion
\mbox{\boldmath{$\xi$}}%
%EndExpansion
}, \label{27c}%
\end{align}
we have%
\begin{align}
1  &  \equiv%
%TCIMACRO{\TeXButton{sigma}{\mbox{\boldmath{$\sigma$}}}}%
%BeginExpansion
\mbox{\boldmath{$\sigma$}}%
%EndExpansion
_{0}=%
%TCIMACRO{\TeXButton{iota}{\mbox{\boldmath{$\iota$}}}}%
%BeginExpansion
\mbox{\boldmath{$\iota$}}%
%EndExpansion
(\mathbf{s}_{1}\otimes\mathbf{s}^{\dot{1}}+\mathbf{s}_{2}\otimes
\mathbf{s}^{\dot{2}}),\nonumber\\%
%TCIMACRO{\TeXButton{sigma}{\mbox{\boldmath{$\sigma$}}}}%
%BeginExpansion
\mbox{\boldmath{$\sigma$}}%
%EndExpansion
_{1}  &  =-%
%TCIMACRO{\TeXButton{iota}{\mbox{\boldmath{$\iota$}}}}%
%BeginExpansion
\mbox{\boldmath{$\iota$}}%
%EndExpansion
(\mathbf{s}_{1}\otimes\mathbf{s}^{\dot{2}}+\mathbf{s}_{2}\otimes
\mathbf{s}^{\dot{1}}),\nonumber\\%
%TCIMACRO{\TeXButton{sigma}{\mbox{\boldmath{$\sigma$}}}}%
%BeginExpansion
\mbox{\boldmath{$\sigma$}}%
%EndExpansion
_{2}  &  =%
%TCIMACRO{\TeXButton{iota}{\mbox{\boldmath{$\iota$}}}}%
%BeginExpansion
\mbox{\boldmath{$\iota$}}%
%EndExpansion
[\mathbf{i}(\mathbf{s}_{1}\otimes\mathbf{s}^{\dot{2}}-\mathbf{s}_{2}%
\otimes\mathbf{s}^{\dot{1}})],\nonumber\\%
%TCIMACRO{\TeXButton{sigma}{\mbox{\boldmath{$\sigma$}}}}%
%BeginExpansion
\mbox{\boldmath{$\sigma$}}%
%EndExpansion
_{3}  &  =-%
%TCIMACRO{\TeXButton{iota}{\mbox{\boldmath{$\iota$}}}}%
%BeginExpansion
\mbox{\boldmath{$\iota$}}%
%EndExpansion
(\mathbf{s}_{1}\otimes\mathbf{s}^{\dot{1}}-\mathbf{s}_{2}\otimes
\mathbf{s}^{\dot{2}}). \label{27d}%
\end{align}

>From this it follows the identification
\begin{equation}
\mathbb{R}_{3,0}\simeq\mathbb{R}_{1,3}^{(0)}\simeq\mathbb{C(}2\mathbb{)=}%
\mathbf{I}\otimes_{\mathbb{C}}\mathbf{\dot{I},} \label{27dd}%
\end{equation}
and then, each Pauli number can be written as an appropriate sum of Clifford
products of algebraic contravariant undotted spinors and algebraic\ covariant
dotted spinors. And, of course, a representative of a Pauli number in
$\mathbb{C}^{2}$ can be written as an appropriate Kronecker product of a
complex column vector \ by a complex row vector.

Take an arbitrary $\mathbf{P\in}\mathbb{R}_{3,0}$ such that
\begin{equation}
\mathbf{P}=\frac{1}{j!}p_{\text{ }}^{_{\mathbf{k}_{1}\mathbf{k}_{2}%
...\mathbf{k}_{j}}}\text{ }%
%TCIMACRO{\TeXButton{sigma}{\mbox{\boldmath{$\sigma$}}}}%
%BeginExpansion
\mbox{\boldmath{$\sigma$}}%
%EndExpansion
_{\mathbf{k}_{1}\mathbf{k}_{2}\mathbf{...k}_{j}}, \label{27e}%
\end{equation}
where $p^{_{\mathbf{k}_{1}\mathbf{k}_{2}...\mathbf{k}_{j}}}\in\mathbb{R}$ and%
\begin{equation}%
%TCIMACRO{\TeXButton{sigma}{\mbox{\boldmath{$\sigma$}}}}%
%BeginExpansion
\mbox{\boldmath{$\sigma$}}%
%EndExpansion
_{_{\mathbf{k}_{1}\mathbf{k}_{2}...\mathbf{k}_{j}}}=%
%TCIMACRO{\TeXButton{sigma}{\mbox{\boldmath{$\sigma$}}}}%
%BeginExpansion
\mbox{\boldmath{$\sigma$}}%
%EndExpansion
_{\mathbf{k}_{1}}...%
%TCIMACRO{\TeXButton{sigma}{\mbox{\boldmath{$\sigma$}}}}%
%BeginExpansion
\mbox{\boldmath{$\sigma$}}%
%EndExpansion
_{\mathbf{k}_{j}},\hspace{0.15in}\text{and }%
%TCIMACRO{\TeXButton{sigma}{\mbox{\boldmath{$\sigma$}}}}%
%BeginExpansion
\mbox{\boldmath{$\sigma$}}%
%EndExpansion
_{0}\equiv1\in\mathbb{R}. \label{27f}%
\end{equation}
\ 

With the identification $\mathbb{R}_{3,0}\simeq\mathbb{R}_{1,3}^{(0)}%
\simeq\mathbf{I}\otimes_{\mathbb{C}}\mathbf{\dot{I}}$, we can also write
\begin{equation}
\mathbf{P=P}_{\text{ \ }\dot{B}}^{A}%
%TCIMACRO{\TeXButton{iota}{\mbox{\boldmath{$\iota$}}}}%
%BeginExpansion
\mbox{\boldmath{$\iota$}}%
%EndExpansion
(\mathbf{s}_{A}\otimes\mathbf{s}^{\dot{B}})=\mathbf{P}_{\text{ \ }\dot{B}}%
^{A}\mathbf{s}_{A}\mathbf{s}^{\dot{B}}, \label{27g}%
\end{equation}
where the $\mathbf{P}_{\text{ \ }\dot{B}}^{A}=\mathbf{X}_{\text{ \ }\dot{B}%
}^{A}+\mathbf{iY}_{\text{ \ }\dot{B}}^{A}$, $\mathbf{X}_{\text{ \ }\dot{B}%
}^{A},\mathbf{Y}_{\text{ \ }\dot{B}}^{A}\in\mathbb{R}$.

Finally, the matrix representative of the Pauli number $\mathbf{P\in
}\mathbb{R}_{3,0}$\textbf{ }is $P\in\mathbb{C(}2)$ given by
\begin{equation}
P=P_{\text{ \ }\dot{B}}^{A}s_{A}s^{\dot{B}}, \label{27h}%
\end{equation}
with $P_{\text{ \ }\dot{B}}^{A}\in\mathbb{C}$ and
\begin{equation}%
\begin{array}
[c]{cc}%
s_{1}=\left(
\begin{array}
[c]{c}%
1\\
0
\end{array}
\right)  & s_{2}=\left(
\begin{array}
[c]{c}%
0\\
1
\end{array}
\right) \\
s^{\dot{1}}=\left(
\begin{array}
[c]{cc}%
1 & 0
\end{array}
\right)  & s^{\dot{2}}=\left(
\begin{array}
[c]{cc}%
0 & 1
\end{array}
\right)  .
\end{array}
\label{27i}%
\end{equation}

It is convenient for our purposes to introduce also covariant undotted spinors
and contravariant dotted spinors. Let $\varphi\in\mathbb{C}^{2}$ be given as
in Eq.(\ref{26}). We define the \textit{covariant} \ version of undotted
spinor $\varphi\in\mathbb{C}^{2}$ as $\varphi^{\ast}$ $\in(\mathbb{C}^{2}%
)^{t}\simeq\mathbb{C}_{2}$ such that%
\begin{align}
\varphi^{\ast}  &  =\left(  \varphi_{1},\varphi_{2}\right)  \equiv\varphi
_{A}s^{A},\nonumber\\
\varphi_{A}  &  =\varphi^{B}\varepsilon_{BA,\hspace{0.15in}}\varphi
^{B}=\varepsilon^{BA}\varphi_{A},\nonumber\\
s^{1}  &  =\left(
\begin{array}
[c]{cc}%
1 & 0
\end{array}
\right)  ,\hspace{0.15in}s^{2}=\left(
\begin{array}
[c]{cc}%
0 & 1
\end{array}
\right)  , \label{27j}%
\end{align}
where\footnote{The symbol \textrm{adiag} means the antidiagonal matrix.}
$\varepsilon_{AB}=\varepsilon^{AB}=\mathrm{adiag}(1,-1)$. We can write due to
the above identifications that there exists $\varepsilon\in\mathbb{C}(2)$
given by Eq.(\ref{27''}) which can be written also as
\begin{equation}
\varepsilon=\varepsilon^{AB}s_{A}\boxtimes s_{B}=\varepsilon_{AB}%
s^{A}\boxtimes s^{B}=\left(
\begin{array}
[c]{cc}%
0 & 1\\
-1 & 0
\end{array}
\right)  =\mathrm{i}\sigma_{2} \label{27K}%
\end{equation}
where $\boxtimes$ denote the \textit{Kronecker} product of matrices. We have,
e.g.,%
\begin{align}
s_{1}\boxtimes s_{2}  &  =\left(
\begin{array}
[c]{c}%
1\\
0
\end{array}
\right)  \boxtimes\left(
\begin{array}
[c]{c}%
0\\
1
\end{array}
\right)  =\left(
\begin{array}
[c]{c}%
1\\
0
\end{array}
\right)  \left(
\begin{array}
[c]{cc}%
0 & 1
\end{array}
\right)  =\left(
\begin{array}
[c]{cc}%
0 & 1\\
0 & 0
\end{array}
\right)  ,\nonumber\\
s^{1}\boxtimes s^{1}  &  =\left(
\begin{array}
[c]{cc}%
1 & 0
\end{array}
\right)  \boxtimes\left(
\begin{array}
[c]{cc}%
0 & 1
\end{array}
\right)  =\left(
\begin{array}
[c]{c}%
1\\
0
\end{array}
\right)  \left(
\begin{array}
[c]{cc}%
1 & 0
\end{array}
\right)  =\left(
\begin{array}
[c]{cc}%
1 & 0\\
0 & 0
\end{array}
\right)  . \label{27l}%
\end{align}

We now introduce the \textit{contravariant }version of the dotted spinor%
\[
\dot{\xi}=\left(
\begin{array}
[c]{cc}%
\xi_{\dot{1}} & \xi_{\dot{2}}%
\end{array}
\right)  \in\mathbb{C}_{2}%
\]
as being $\dot{\xi}^{\ast}\in\mathbb{C}^{2}$ such that%
\begin{align}
\dot{\xi}^{\ast}  &  =\left(
\begin{array}
[c]{c}%
\xi^{\dot{1}}\\
\xi^{\dot{2}}%
\end{array}
\right)  =\xi^{\dot{A}}s_{\dot{A}},\nonumber\\
\xi^{\dot{B}}  &  =\varepsilon^{\dot{B}\dot{A}}\xi_{\dot{A}},\hspace
{0.15in}\xi_{\dot{A}}=\varepsilon_{\dot{B}\dot{A}}\text{ }\xi^{\dot{B}%
},\nonumber\\
s_{\dot{1}}  &  =\left(
\begin{array}
[c]{c}%
1\\
0
\end{array}
\right)  ,s_{\dot{2}}=\left(
\begin{array}
[c]{c}%
0\\
1
\end{array}
\right)  , \label{27m}%
\end{align}
where \ $\varepsilon_{\dot{A}\dot{B}}=\varepsilon^{\dot{A}\dot{B}%
}=\mathrm{adiag}(1,-1)$. Then, due to the above identifications we see that
there exists $\dot{\varepsilon}\in\mathbb{C}(2)$ such that
\begin{equation}
\dot{\varepsilon}=\varepsilon^{\dot{A}\dot{B}}s_{\dot{A}}\boxtimes s_{\dot{B}%
}=\varepsilon_{\dot{A}\dot{B}}s^{\dot{A}}\boxtimes\dot{s}^{B}=\left(
\begin{array}
[c]{cc}%
0 & 1\\
-1 & 0
\end{array}
\right)  =\varepsilon. \label{27n}%
\end{equation}

Also, recall that even if \ $\{\mathbf{s}_{A}\}$,$\{\mathbf{s}_{\dot{A}}\}$
and $\{s^{\dot{A}}\}$,$\{s^{A}\}$ are bases of distinct spaces, we can
identify their matrix representations, as it is obvious from the above
formulas. So, we have $s_{A}\equiv s_{\dot{A}}$ and also $s^{\dot{A}}=s^{A}$.
This is the reason for the representation of a dotted covariant spinor as in
Eq.(\ref{27'}). Moreover, the above identifications permit us to write the
\textit{matrix} \textit{representation} of a Pauli number $\mathbf{P\in
}\mathbb{R}_{3,0}$ as, e.g.,
\begin{equation}
P=P_{AB}s^{A}\boxtimes s^{B} \label{27o}%
\end{equation}
besides the representation given by Eq.(\ref{27h}).

\section{Clifford and Spinor Bundles}

\subsection{Preliminaries}

To characterize in a rigorous mathematical way the \textit{basic} field
variables used in M. Sachs `unified' field theory \cite{s2,s3,s4}, we
shall\ need to recall some results of the theory of spinor fields on
Lorentzian spacetimes. Here we follow the approach given in \cite{28,moro}%
.\footnote{Another important reference on the subject of spinor fields (in the
spirit of this \ work) is \cite{lami}, which however only deals with the case
of spinor fields on Riemannian manifolds.}

Recall that a Lorentzian manifold is a pair $(M,g)$, where $g\in\sec T^{2,0}M$
is a Lorentzian metric of signature $(1,3)$, i.e., for all $x\in M$,
$T_{x}M\simeq T_{x}^{\ast}M\simeq\mathbb{R}^{1,3}$, where $\mathbb{R}^{1,3}$
is the vector Minkowski space.

Recall that a Lorentzian spacetime is a pentuple $(M,g,D\mathbf{,\tau}%
_{g},\mathbf{\uparrow})$ where $(M,g,$\linebreak$\mathbf{\tau}_{g})$ is an
oriented Lorentzian manifold\footnote{Oriented by the volume element
$\mathbf{\tau}_{g}\in\sec%
%TCIMACRO{\dbigwedge \nolimits^{4}}%
%BeginExpansion
{\displaystyle\bigwedge\nolimits^{4}}
%EndExpansion
T^{\ast}M$.} \ which is also time oriented by an appropriated equivalence
relation\footnote{See \cite{sw} for details.} (denoted $\uparrow$) for the
timelike vectors at the tangent space $T_{x}M$, $\forall x\in M$.
$D$\textbf{\ }is a linear connection for $M$ such that $Dg=0$, $\mathbf{\Theta
}(D)=0$, $\mathcal{R}(D)\neq0$, where $\mathbf{\Theta}$ and $\mathcal{R}$ are
respectively the torsion and curvature tensors of $D$.

Now, M. Sachs theory as described in \cite{s2,s3,s4} uses spinor fields. These
objects are sections of so-called \textit{spinor bundles}, which only exist in
\textit{spin manifolds}. The ones used in Sachs theory are the matrix
representation of sections of the bundles of dotted spinor fields, i.e.,
$S(M)=P_{\mathrm{Spin}_{1,3}^{e}}(M)\times_{D^{(\frac{1}{2},0)}}\mathbb{C}%
^{2}$ and the matrix representation of the bundle of undotted spinor fields,
here denoted by $\bar{S}(M)=P_{\mathrm{Spin}_{1,3}^{e}}(M)\times
_{D^{(0,\frac{1}{2})}}\mathbb{C}_{2}$ . In the previous formula $D^{(\frac
{1}{2},0)}$ and $D^{(0,\frac{1)}{2}}$ \ are the two fundamental non equivalent
$2$-dimensional representations of $Sl(2,\mathbb{C)\simeq}\mathrm{Spin}%
_{1,3}^{e}$, the universal covering group of $\mathrm{SO}_{1,3}^{e}$, the
restrict orthochronous Lorentz group. $P_{\mathrm{Spin}_{1,3}^{e}}(M)$ is a
principal bundle called the spin structure bundle\footnote{It is a covering
space of $P_{\mathrm{SO}_{1,3}^{e}}(M)$. See, e.g., \cite{moro} for details. A
section of \ $P_{\mathrm{Spin}_{1,3}^{e}}(M)$ is called a spin frame, which
can be identified as pair $(\Sigma,u)$ where for any $x\in M$, $\Sigma(x)$ is
an othonormal frame and $u(x)$ belongs to the $\mathrm{Spin}_{1,3}^{e}$.}. We
recall that it is a classical result \ (Geroch theorem \cite{g1}) that a
$4$-dimensional Lorentzian manifold is a spin manifold if and only if
$P_{\mathrm{SO}_{1,3}^{e}}(M)$ has a global section\footnote{In what follows
$P_{\mathrm{SO}_{1,3}^{e}}(M)$ denotes the principal bundle of oriented
\textit{Lorentz tetrads}. We presuppose that the reader is acquainted with the
structure of $P_{\mathrm{SO}_{1,3}^{e}}(M)$, whose sections are the time
oriented and oriented orthonormal frames.}, i.e., if there exists a set
$\{\mathbf{e}_{\mathbf{0}},\mathbf{e}_{\mathbf{1}},\mathbf{e}_{\mathbf{2}%
},\mathbf{e}_{\mathbf{3}}\}$ of orthonormal fields defined for all $x\in M$.
\ In other word, for spinor fields to exist in a $4$-dimensional spacetime the
orthonormal frame bundle must be \textit{trivial.}

Now, the so-called tangent ($TM$) and cotangent ($T^{\ast}M$) bundles, the
tensor bundle ($\oplus_{r,s}\otimes_{s}^{r}TM)$ and the bundle of differential
forms for the spacetime are the bundles denoted by
\begin{align}
TM  &  =P_{\mathrm{SO}_{1,3}^{e}}(M)\times_{\rho_{_{1,3}}}\mathbb{R}%
^{1,3},\hspace{0.15in}T^{\ast}M=P_{\mathrm{SO}_{1,3}^{e}}(M)\times_{\rho
_{1,3}^{\ast}}\mathbb{R}^{1,3},\label{S1}\\
\oplus_{r,s}\otimes_{s}^{r}TM  &  =P_{\mathrm{SO}_{1,3}^{e}}(M)\times
_{\otimes_{s}^{r}\rho_{_{1,3}}}\mathbb{R}^{1,3},\hspace{0.15in}%
%TCIMACRO{\dbigwedge }%
%BeginExpansion
{\displaystyle\bigwedge}
%EndExpansion
T^{\ast}M=P_{\mathrm{SO}_{1,3}^{e}}(M)\times_{\Lambda_{\rho_{1,3}^{\ast}}^{k}}%
%TCIMACRO{\dbigwedge }%
%BeginExpansion
{\displaystyle\bigwedge}
%EndExpansion
\mathbb{R}^{1,3}.\nonumber
\end{align}

In Eqs.(\ref{S1})
\begin{equation}
\rho_{_{1,3}}:\mathrm{SO}_{1,3}^{e}\rightarrow\mathrm{SO}^{e}(\mathbb{R}%
^{1,3}) \label{S2}%
\end{equation}
\ is the standard vector representation of $\mathrm{SO}_{1,3}^{e}$ usually
denoted by \footnote{See, e.g., \cite{miller} if you need details.}
$D^{(\frac{1}{2},\frac{1}{2})}=$ $D^{(\frac{1}{2},0)}\otimes D^{\left(
0,\frac{1}{2}\right)  }$and $\rho_{1,3}^{\ast}$ is the dual (vector)
representation $\rho_{1,3}^{\ast}\left(  l)=\rho_{1,3}(l^{-1}\right)  ^{t}$.
\ Also $\otimes_{s}^{r}\rho_{_{1,3}}$ and $\Lambda_{\rho_{1,3}^{\ast}}^{k}$
are the induced tensor product and induced exterior power product
representations of $\mathrm{SO}_{1,3}^{e}$. We now briefly recall the
definition and some properties of the Clifford bundle of multivector fields
\cite{28}. We have,%
\begin{align}
\mathcal{C\ell}(TM)  &  =P_{\mathrm{SO}_{1,3}^{e}}(M)\times_{c\ell
_{\rho_{_{1,3}}}}\mathbb{R}_{1,3}\nonumber\\
&  =P_{\mathrm{Spin}_{1,3}^{e}}(M)\times_{\mathrm{Ad}}\mathbb{R}_{1,3}.
\label{C3}%
\end{align}
Now, recall that \cite{lounesto} $\mathrm{Spin}_{1,3}^{e}\subset
\mathbb{R}_{1,3}^{(0)}$. Consider the $2$-$1$ homomorphism $\mathrm{h}%
:\mathrm{Spin}_{1,3}^{e}\rightarrow\mathrm{SO}_{1,3}^{e},\mathrm{h}(\pm u)=l$.
Then $c\ell_{_{\rho_{_{1,3}}}}$ is the following representation of
$\mathrm{SO}_{1,3}^{e}$,
\begin{align}
c\ell_{_{\rho_{_{1,3}}}}  &  :\mathrm{SO}_{1,3}^{e}\rightarrow\mathrm{Aut}%
(\mathbb{R}_{1,3}),\nonumber\\
c\ell_{_{\rho_{_{1,3}}}}(L)  &  =\mathrm{Ad}_{u}:\mathbb{R}_{1,3}%
\rightarrow\mathbb{R}_{1,3},\nonumber\\
\mathrm{Ad}_{u}(\mathbf{m)}  &  =u\mathbf{m}u^{-1} \label{C.4}%
\end{align}
i.e., it\ is the standard orthogonal transformation of $\mathbb{R}_{1,3}$
induced by an orthogonal transformation of $\mathbb{R}^{1,3}$. Note that
$\mathrm{Ad}_{u}$ act on vectors as the $D^{(\frac{1}{2},\frac{1}{2})}$
representation of $\mathrm{SO}_{1,3}^{e}$ and on multivectors as the induced
exterior power representation of that group. Indeed, observe, e.g., that for
$\mathbf{v\in}\mathbb{R}^{1,3}\subset\mathbb{R}_{1,3}$ we have in standard
notation
\[
L\mathbf{v=v}^{\nu}L_{\nu}^{\mu}\mathbf{m}_{\mathbf{\mu}}=\mathbf{v}^{\nu
}u\mathbf{m}_{\nu}u^{-1}=u\mathbf{v}u^{-1}.
\]

The proof of the second line of Eq.(\ref{C3}) is as follows. Consider the
representation%
\begin{align}
\mathrm{Ad}  &  :\mathrm{Spin}_{1,3}^{e}\rightarrow\mathrm{Aut}(\mathbb{R}%
_{1,3}),\nonumber\\
\mathrm{Ad}_{u}  &  :\mathbb{R}_{1,3}\rightarrow\mathbb{R}_{1,3}%
,\hspace{0.15in}\mathrm{Ad}_{u}\left(  m\right)  =umu^{-1}. \label{C5}%
\end{align}

Since $\mathrm{Ad}_{-1}=1$($=$ \textrm{identity) }the representation
$\mathrm{Ad}$ descends to a representation of $\mathrm{SO}_{1,3}^{e}$. This
representation is just $c\ell(\rho_{_{1,3}})$, from where the desired result follows.

Sections of $\mathcal{C\ell}(TM)$ can be called Clifford fields (of
multivectors). The sections of the even subbundle $\mathcal{C\ell}%
^{(0)}(TM)=P_{\mathrm{Spin}_{1,3}^{e}}(M)\times_{\mathrm{Ad}}\mathbb{R}%
_{1,3}^{(0)}$ may be called Pauli fields (of multivectors). Define the real
spinor bundles%

\begin{equation}
\mathcal{S}(M)=P_{\mathrm{Spin}_{1,3}^{e}}(M)\times_{l}\mathbf{I,\hspace
{0.15in}}\mathcal{\dot{S}}(M)=P_{\mathrm{Spin}_{1,3}^{e}}(M)\times
_{r}\mathbf{\dot{I}} \label{C.5a}%
\end{equation}
where $l$ stands for a left modular representation of $\mathrm{Spin}_{1,3}%
^{e}$ in $\mathbb{R}_{1,3}$ that mimics the $D^{(\frac{1}{2},0)}$
representation of $Sl(2,\mathbb{C)}$ and $r$ stands for a right modular
representation of $\mathrm{Spin}_{1,3}^{e}$ in $\mathbb{R}_{1,3}$ that mimics
the $D^{(0,\frac{1}{2})}$ representation of $Sl(2,\mathbb{C)}$.

Also recall that if $\bar{S}(M)$ is the bundle whose sections are the spinor
fields $\bar{\varphi}=(\bar{\varphi}_{1},\bar{\varphi}_{2})=\dot{\varphi
}\varepsilon=(\varphi^{\dot{1}},\varphi^{\dot{2}})$, then it is isomorphic to
the space of contravariant dotted spinors. We have,%
\begin{equation}
S(M)\mathbf{\simeq}P_{\mathrm{Spin}_{1,3}^{e}}(M)\times_{D^{(\frac{1}{2},0)}%
}\mathbb{C}^{2},\mathbf{\hspace{0.15in}}\dot{S}\left(  M\right)
\mathbf{\simeq}P_{\mathrm{Spin}_{1,3}^{e}}(M)\times_{D^{(0,\frac{1}{2})}%
}\mathbb{C}_{2}\simeq\bar{S}(M), \label{C6}%
\end{equation}
and from our playing with the Pauli algebra and dotted and undotted spinors in
section 2 we have that:
\begin{equation}
\mathcal{S}(M)\simeq S(M),\hspace{0.15in}\mathcal{\dot{S}}(M)\simeq\dot
{S}\left(  M\right)  \mathbf{\simeq}\bar{S}(M). \label{C.6bis}%
\end{equation}

Then, we have the obvious isomorphism%
\begin{align}
\mathcal{C\ell}^{(0)}(TM)  &  =P_{\mathrm{Spin}_{1,3}^{e}}(M)\times
_{\mathrm{Ad}}\mathbb{R}_{1,3}^{(0)}\nonumber\\
&  =P_{\mathrm{Spin}_{1,3}^{e}}(M)\times_{l\otimes r}\mathbf{I\otimes
}_{\mathbb{C}}\mathbf{\dot{I}}\nonumber\\
&  =\mathcal{S}(M)\otimes_{\mathbb{C}}\mathcal{\dot{S}}(M). \label{C7}%
\end{align}

Let us now introduce the following (complex) bundle,%
\begin{equation}
\mathbb{C}\ell^{(0)}(M)=P_{\mathrm{Spin}_{1,3}^{e}}(M)\times_{_{D^{(\frac
{1}{2}0)}\otimes D^{(0,\frac{1}{2})}}}\mathbb{C}(2). \label{C8'}%
\end{equation}
It is clear that%
\begin{equation}
\mathbb{C}\ell^{(0)}(M)=S(M)\otimes_{\mathbb{C}}\bar{S}(M)\simeq
\mathcal{C\ell}^{(0)}(M). \label{C8}%
\end{equation}

Finally, we consider the bundle
\begin{equation}
\mathcal{C\ell}^{(0)}(TM)\otimes%
%TCIMACRO{\dbigwedge }%
%BeginExpansion
{\displaystyle\bigwedge}
%EndExpansion
T^{\ast}M\simeq\mathbb{C}\ell^{(0)}(M)\otimes%
%TCIMACRO{\dbigwedge }%
%BeginExpansion
{\displaystyle\bigwedge}
%EndExpansion
T^{\ast}M. \label{C9}%
\end{equation}

Sections of $\mathcal{C\ell}^{(0)}(TM)\otimes%
%TCIMACRO{\dbigwedge }%
%BeginExpansion
{\displaystyle\bigwedge}
%EndExpansion
T^{\ast}M$ may be called \textit{Pauli valued differential forms} and sections
of $\mathbb{C}\ell^{(0)}(M)\otimes%
%TCIMACRO{\dbigwedge }%
%BeginExpansion
{\displaystyle\bigwedge}
%EndExpansion
T^{\ast}M$ may be called \textit{matrix Pauli valued differential
forms\footnote{A detailed theory of Clifford valued differential forms is
given in \cite{cliforms}.}}. \textrm{ }

\bigskip\ Denote by $\mathcal{C\ell}_{(0,2)}^{\left(  0\right)  }(TM)$ $\ $the
seven dimensional subbundle $\left(  \mathbb{R\oplus}%
%TCIMACRO{\dbigwedge \nolimits^{2}}%
%BeginExpansion
{\displaystyle\bigwedge\nolimits^{2}}
%EndExpansion
TM\right)  \subset%
%TCIMACRO{\dbigwedge }%
%BeginExpansion
{\displaystyle\bigwedge}
%EndExpansion
TM\hookrightarrow\mathcal{C\ell}^{(0)}(TM)\subset\mathcal{C\ell}(TM)$. \ Now,
\ let $\langle x^{\mu}\rangle$ be the coordinate functions of a chart of the
maximal atlas of $M$. The fundamental field variable of Sachs theory can be
described as%
\[
\mathbf{Q=q}_{\mu}\otimes dx^{\mu}\equiv\mathbf{q}_{\mu}dx^{\mu}\mathbf{\in
}\sec\mathcal{C\ell}_{(0,2)}^{(0)}(TM)\otimes%
%TCIMACRO{\dbigwedge }%
%BeginExpansion
{\displaystyle\bigwedge}
%EndExpansion
T^{\ast}M\subset\sec\mathcal{C\ell}^{(0)}(TM)\otimes%
%TCIMACRO{\dbigwedge }%
%BeginExpansion
{\displaystyle\bigwedge}
%EndExpansion
T^{\ast}M
\]
i.e., a Pauli valued $1$-form obeying certain conditions to be presented
below. If we work (as Sachs did) with $\mathbb{C}\ell^{(0)}(M)\otimes%
%TCIMACRO{\dbigwedge }%
%BeginExpansion
{\displaystyle\bigwedge}
%EndExpansion
T^{\ast}M$, a representative of $\mathbf{Q}$ is $Q\in\sec\mathbb{C}\ell
^{(0)}(M)\otimes%
%TCIMACRO{\dbigwedge }%
%BeginExpansion
{\displaystyle\bigwedge}
%EndExpansion
T^{\ast}M$ such that\footnote{Note that a bold index (sub or superscript), say
$\mathbf{a}$ take the values $0,1,2,3$.}%
\begin{equation}
Q=q_{\mu}(x)dx^{\mu}=h_{\mu}^{\mathbf{a}}(x)dx^{\mu}\sigma_{\mathbf{a}},
\label{S4}%
\end{equation}
where $\sigma_{0}=\left(
\begin{array}
[c]{cc}%
1 & 0\\
0 & 1
\end{array}
\right)  $ and $\sigma_{j}$ ($j\mathbf{=}1,2,3$) are the Pauli matrices. \ We
observe that the notation anticipates the fact that in Sachs theory the
variables $h_{\mu}^{\mathbf{a}}(x)$ define the set $\{\theta^{\mathbf{a}%
}\}\equiv\{\theta^{\mathbf{0}},\theta^{\mathbf{1}},\theta^{\mathbf{2}}%
,\theta^{\mathbf{3}}\}$ with
\begin{equation}
\theta^{\mathbf{a}}=h_{\mu}^{\mathbf{a}}dx^{\mu}\in\sec%
%TCIMACRO{\dbigwedge }%
%BeginExpansion
{\displaystyle\bigwedge}
%EndExpansion
T^{\ast}M, \label{S5}%
\end{equation}
which is the dual basis of $\ \{\mathbf{e}_{\mathbf{a}}\}\equiv\{\mathbf{e}%
_{\mathbf{0}},\mathbf{e}_{\mathbf{1}},\mathbf{e}_{\mathbf{2}},\mathbf{e}%
_{\mathbf{3}}\}$, $\mathbf{e}_{\mathbf{a}}\in\sec TM$. \ We denote by
$\{e_{\mu}\}=\{e_{0},e_{1},e_{2},e_{3}\}$, a coordinate basis associated with
the local chart $\langle x^{\mu}\rangle$ covering $U\subset M$ . We have
\ $e_{\mu}=h_{\mu}^{\mathbf{a}}\mathbf{e}_{\mathbf{a}}\in\sec TM$, \ and the
set $\{e_{\mu}\}$ is the dual basis of $\{dx^{\mu}\}\equiv\{dx^{0}%
,dx^{1},dx^{2},dx^{3}\}$. We will also use the \textit{reciprocal basis} to a
given basis $\{\mathbf{e}_{\mathbf{a}}\}$, i.e., the set $\{\mathbf{e}%
^{\mathbf{a}}\}\equiv\{\mathbf{e}^{\mathbf{0}},\mathbf{e}^{\mathbf{1}%
},\mathbf{e}^{\mathbf{2}},\mathbf{e}^{\mathbf{3}}\},\mathbf{e}^{\mathbf{a}}%
\in\sec TM$, with $g(\mathbf{e}_{\mathbf{a}},\mathbf{e}^{\mathbf{b}}%
)=\delta_{\mathbf{a}}^{\mathbf{b}}$ \ and the \textit{reciprocal basis} to
$\{\theta^{\mathbf{a}}\}$, i.e., the set\ $\{\theta_{\mathbf{a}}%
\}=\{\theta_{\mathbf{0}},\theta_{\mathbf{1}},\theta_{\mathbf{2}}%
,\theta_{\mathbf{3}}\}$, with $\theta_{\mathbf{a}}(e^{\mathbf{b}})=\delta
_{a}^{\mathbf{b}}$. Recall that since $\eta_{\mathbf{ab}}=g(\mathbf{e}%
_{\mathbf{a}},\mathbf{e}_{\mathbf{b}})$ ,\ we have%

\begin{equation}
g_{\mu\nu}=g\left(  e_{\mu},e_{\nu}\right)  =h_{\mu}^{\mathbf{a}}h_{\nu
}^{\mathbf{b}}\eta_{\mathbf{ab}}. \label{28new}%
\end{equation}

To continue, we define
\begin{equation}
\check{\sigma}_{0}=-\sigma_{0}\text{ and }\check{\sigma}_{\mathbf{j}}%
=\sigma_{\mathbf{j}},\mathbf{j}=1,2,3 \label{28newbis}%
\end{equation}
and
\begin{equation}
\check{Q}=\check{q}_{\mu}(x)dx^{\mu}=h_{\mu}^{\mathbf{a}}(x)dx^{\mu}%
\check{\sigma}_{\mathbf{a}}. \label{29}%
\end{equation}

We note that
\begin{equation}
\sigma_{\mathbf{a}}\check{\sigma}_{\mathbf{b}}+\sigma_{\mathbf{b}}%
\check{\sigma}_{\mathbf{a}}=-2\eta_{\mathbf{ab}}. \label{30}%
\end{equation}

Readers of Sachs' books \cite{s1,s3} will recall that he said that $Q$ is a
representative of a \textit{quaternion}.\footnote{Note that Sachs represented
$Q$ by $d\mathsf{S}$, which is a very dangerous notation, which we avoid.
Sachs notation has lead him in the past \cite{s0} to identifiy $d\mathsf{S}$
with the element of arc of a curve in a Lorentzain manifold, thus producing
unfortunately a lot of misunderstandings,as showed in \cite{rosa}. On this
issue see also the erronous Sachs reply to \cite{rosa} in \cite{s4}. See also
\cite{roli}.} From our previous discussion we see that this statement is
\textit{not }correct.\footnote{Nevertheless most of the calculations done by
Sachs in \cite{s1,s3} are correct because he worked always with the matrix
representation of $\mathbf{Q}$. However, his claim of having produce an
unified field theory of gravitation and electromagnetism is wrong as we shall
prove in a following paper\cite{cliforms}.} Sachs identification is a
dangerous one, because the quaternions close a division algebra, also-called a
noncommutative field or skew-field and objects like $\mathbf{Q=q}_{\mu}\otimes
dx^{\mu}\mathbf{\in}\sec\mathcal{C\ell}_{(0,2)}^{(0)}(TM)\otimes%
%TCIMACRO{\dbigwedge }%
%BeginExpansion
{\displaystyle\bigwedge}
%EndExpansion
T^{\ast}M\subset\sec\mathcal{C\ell}^{(0)}(TM)\otimes%
%TCIMACRO{\dbigwedge }%
%BeginExpansion
{\displaystyle\bigwedge}
%EndExpansion
T^{\ast}M$, called \textit{paravector} fields, did not close a
\textit{division} algebra.

Next we introduce a tensor product of sections $\mathbf{A,B}\in\sec
\mathcal{C\ell}^{(0)}(M)\otimes%
%TCIMACRO{\dbigwedge }%
%BeginExpansion
{\displaystyle\bigwedge}
%EndExpansion
T^{\ast}M$. Before we do that we recall that from now on%
\begin{equation}
\{1,%
%TCIMACRO{\TeXButton{sigma}{\mbox{\boldmath{$\sigma$}}}}%
%BeginExpansion
\mbox{\boldmath{$\sigma$}}%
%EndExpansion
_{\mathbf{k}},%
%TCIMACRO{\TeXButton{sigma}{\mbox{\boldmath{$\sigma$}}}}%
%BeginExpansion
\mbox{\boldmath{$\sigma$}}%
%EndExpansion
_{\mathbf{k}_{1}\mathbf{k}_{2}},\text{\textsl{i}}=%
%TCIMACRO{\TeXButton{sigma}{\mbox{\boldmath{$\sigma$}}}}%
%BeginExpansion
\mbox{\boldmath{$\sigma$}}%
%EndExpansion
_{\mathbf{123}}\}, \label{30'}%
\end{equation}
refers to a basis of $\mathcal{C\ell}^{(0)}(M)$, i.e., they are
fields.\footnote{We hope that in using (for symbol economy) the same notation
as in section 2 where the $\{1,%
%TCIMACRO{\TeXButton{sigma}{\mbox{\boldmath{$\sigma$}}}}%
%BeginExpansion
\mbox{\boldmath{$\sigma$}}%
%EndExpansion
_{\mathbf{k}},%
%TCIMACRO{\TeXButton{sigma}{\mbox{\boldmath{$\sigma$}}}}%
%BeginExpansion
\mbox{\boldmath{$\sigma$}}%
%EndExpansion
_{\mathbf{k}_{1}\mathbf{k}_{2}},%
%TCIMACRO{\TeXButton{sigma}{\mbox{\boldmath{$\sigma$}}}}%
%BeginExpansion
\mbox{\boldmath{$\sigma$}}%
%EndExpansion
_{\mathbf{123}}\}$ is a basis of $\mathbb{R}_{1,3}^{(0)}\simeq\mathbb{R}%
_{3.0}$ will produce no confusion.}

Recalling Eq.(\ref{27f}) we introduce the (obvious) notation
\begin{equation}
\mathbf{A}=\frac{1}{j!}a_{\mu}^{_{\mathbf{k}_{1}\mathbf{k}_{2}...\mathbf{k}%
_{j}}}\text{ }%
%TCIMACRO{\TeXButton{sigma}{\mbox{\boldmath{$\sigma$}}}}%
%BeginExpansion
\mbox{\boldmath{$\sigma$}}%
%EndExpansion
_{\mathbf{k}_{1}\mathbf{k}_{2}...\mathbf{k}_{j}}dx^{\mu},\hspace
{0.15in}\mathbf{B}=\frac{1}{l!}b_{\mu}^{_{\mathbf{k}_{1}\mathbf{k}%
_{2}...\mathbf{k}_{l}}}%
%TCIMACRO{\TeXButton{sigma}{\mbox{\boldmath{$\sigma$}}}}%
%BeginExpansion
\mbox{\boldmath{$\sigma$}}%
%EndExpansion
_{\mathbf{k}_{1}\mathbf{k}_{2}...\mathbf{k}_{l}}dx^{\mu}, \label{31}%
\end{equation}
where the $a_{\mu}^{_{\mathbf{k}_{1}\mathbf{k}_{2}...\mathbf{k}_{j}}},b_{\mu
}^{_{\mathbf{k}_{1}\mathbf{k}_{2}...\mathbf{k}_{j}}}$ are, in general,
\textit{real }scalar functions. Then, we define%
\begin{equation}
\mathbf{A}\otimes\mathbf{B}=\frac{1}{j!l!}a_{\mu}^{_{\mathbf{k}_{1}%
\mathbf{k}_{2}...\mathbf{k}_{j}}}b_{\nu}^{_{\mathbf{p}_{1}\mathbf{p}%
_{2}...\mathbf{p}_{l}}}%
%TCIMACRO{\TeXButton{sigma}{\mbox{\boldmath{$\sigma$}}}}%
%BeginExpansion
\mbox{\boldmath{$\sigma$}}%
%EndExpansion
_{\mathbf{k}_{1}\mathbf{k}_{2}...\mathbf{k}_{j}}%
%TCIMACRO{\TeXButton{sigma}{\mbox{\boldmath{$\sigma$}}}}%
%BeginExpansion
\mbox{\boldmath{$\sigma$}}%
%EndExpansion
_{\mathbf{p}_{1}\mathbf{p}_{2}...\mathbf{p}_{l}}dx^{\mu}\otimes dx^{\nu}.
\label{32}%
\end{equation}

Let us now compute the tensor product of $\mathbf{Q}\otimes\mathbf{\check{Q}}$
where $\mathbf{Q}\in\sec\mathcal{C\ell}_{(0,2)}^{(0)}(M)\otimes%
%TCIMACRO{\dbigwedge }%
%BeginExpansion
{\displaystyle\bigwedge}
%EndExpansion
T^{\ast}M$. \ We have,%

\begin{align}
\mathbf{Q}\otimes\mathbf{\check{Q}}  &  =\mathbf{q}_{\mu}(x)dx^{\mu}%
\otimes\mathbf{\check{q}}_{\nu}(x)dx^{v}=\mathbf{q}_{\mu}(x)\mathbf{\check{q}%
}_{\nu}(x)dx^{\mu}\otimes dx^{\nu}\nonumber\\
&  =\mathbf{q}_{\mu}(x)\mathbf{\check{q}}_{\nu}(x)\frac{1}{2}(dx^{\mu}\otimes
dx^{\nu}+dx^{\nu}\otimes dx^{\mu})\nonumber\\
&  +\frac{1}{2}\mathbf{q}_{\mu}(x)\mathbf{\check{q}}_{\nu}(x)(dx^{\mu}\otimes
dx^{\nu}-dx^{\nu}\otimes dx^{\mu})\nonumber\\
&  =\frac{1}{2}(\mathbf{q}_{\mu}(x)\mathbf{\check{q}}_{\nu}(x)+\mathbf{q}%
_{\nu}(x)\mathbf{\check{q}}_{\mu}(x))dx^{\mu}\otimes dx^{\nu}\nonumber\\
&  +\frac{1}{2}\mathbf{q}_{\mu}(x)\mathbf{\check{q}}_{\nu}(x)dx^{\mu}\wedge
dx^{\nu}\label{33}\\
&  =(-g_{\mu\nu}%
%TCIMACRO{\TeXButton{sigma}{\mbox{\boldmath{$\sigma$}}}}%
%BeginExpansion
\mbox{\boldmath{$\sigma$}}%
%EndExpansion
_{0})dx^{\mu}\otimes dx^{\nu}\nonumber\\
&  +\frac{1}{4}(\mathbf{q}_{\mu}(x)\mathbf{\check{q}}_{\nu}(x)-\mathbf{q}%
_{\nu}(x)\mathbf{\check{q}}_{\mu}(x))dx^{\mu}\wedge dx^{\nu}\nonumber\\
&  =-g_{\mu\nu}dx^{\mu}\otimes dx^{\nu}+\frac{1}{2}\mathbf{F}_{\mu\nu}%
^{\prime}dx^{\mu}\wedge dx^{\nu}.\nonumber
\end{align}

In writing Eq.(\ref{33}) we have used $dx^{\mu}\wedge dx^{\nu}\equiv dx^{\mu
}\otimes dx^{\nu}-dx^{\nu}\otimes dx^{\mu}$. Also, using
\begin{align}
g_{\mu\nu}  &  =\eta_{\mathbf{ab}}h_{\mu}^{\mathbf{a}}(x)h_{\nu}^{\mathbf{b}%
}(x),\hspace{0.15in}g=g_{\mu\nu}dx^{\mu}\otimes dx^{v}=\eta_{\mathbf{ab}%
}\theta^{\mathbf{a}}\otimes\theta^{\mathbf{b}}\nonumber\\
\mathbf{F}_{\mu\nu}^{\prime}  &  =\mathbf{F}_{\mu\nu}^{\prime k}%
\text{\textsl{i}}%
%TCIMACRO{\TeXButton{sigma}{\mbox{\boldmath$\sigma$}}}%
%BeginExpansion
\mbox{\boldmath$\sigma$}%
%EndExpansion
_{k}\mathbf{=-}\frac{1}{2}(\varepsilon_{i\text{ }j}^{k}h_{\mu}^{i}(x)h_{\nu
}^{j}(x)\text{ })\text{\textsl{i}}%
%TCIMACRO{\TeXButton{sigma}{\mbox{\boldmath$\sigma$}}}%
%BeginExpansion
\mbox{\boldmath$\sigma$}%
%EndExpansion
_{k};\hspace{0.15in}i,j,k=1,2,3,\nonumber\\
\mathbf{F}^{\prime}  &  =\frac{1}{2}\mathbf{F}_{\mu\nu}^{\prime}dx^{\mu}\wedge
dx^{\nu}=\frac{1}{2}(\mathbf{F}_{\text{ }\mu\nu}^{\prime ij}%
%TCIMACRO{\TeXButton{sigma}{\mbox{\boldmath$\sigma$}}}%
%BeginExpansion
\mbox{\boldmath$\sigma$}%
%EndExpansion
_{i}%
%TCIMACRO{\TeXButton{sigma}{\mbox{\boldmath$\sigma$}}}%
%BeginExpansion
\mbox{\boldmath$\sigma$}%
%EndExpansion
_{j})dx^{\mu}\wedge dx^{\nu}=(\frac{1}{2}\mathbf{F}_{\mu\nu}^{\prime
k}\text{\textsl{i}}%
%TCIMACRO{\TeXButton{sigma}{\mbox{\boldmath$\sigma$}}}%
%BeginExpansion
\mbox{\boldmath$\sigma$}%
%EndExpansion
_{k})dx^{\mu}\wedge dx^{v}\nonumber\\
&  =-\varepsilon_{i\text{ }j}^{k}h_{\mu}^{i}(x)h_{\nu}^{j}(x)\text{ }dx^{\mu
}\wedge dx^{\nu}\text{\textsl{i}}\mathbf{\sigma}_{k}\in\sec%
%TCIMACRO{\dbigwedge \nolimits^{2}}%
%BeginExpansion
{\displaystyle\bigwedge\nolimits^{2}}
%EndExpansion
T^{\ast}M\otimes\mathcal{C\ell}_{(2)}^{(0)}\left(  M\right)  \label{33new}%
\end{align}
we can write Eq.(\ref{33}) as
\begin{align}
\mathbf{Q}\otimes\mathbf{\check{Q}}  &  \mathbf{=\mathbf{Q}}\overset
{s}{\mathbf{\otimes}}\mathbf{\check{Q}\mathbf{+}Q\wedge\check{Q}}\nonumber\\
&  =-g+\mathbf{F.} \label{33bis}%
\end{align}

We can also write%
\begin{equation}
\mathbf{Q}\otimes\mathbf{\check{Q}=-}\eta_{\mathbf{ab}}%
%TCIMACRO{\TeXButton{sigma}{\mbox{\boldmath{$\sigma$}}}}%
%BeginExpansion
\mbox{\boldmath{$\sigma$}}%
%EndExpansion
_{0}\theta^{\mathbf{a}}\otimes\theta^{\mathbf{b}}+\varepsilon_{i\text{ }j}%
^{k}\text{\textsl{i}}%
%TCIMACRO{\TeXButton{sigma}{\mbox{\boldmath$\sigma$}}}%
%BeginExpansion
\mbox{\boldmath$\sigma$}%
%EndExpansion
_{k}\theta^{i}\wedge\theta^{j}. \label{33biss}%
\end{equation}

The above formulas show very clearly the mathematical nature of $\mathbf{F}$,
it is a $2$-form with values on the subspace of multivector Clifford fields,
i.e., $\mathbf{F:}%
%TCIMACRO{\dbigwedge \nolimits^{2}}%
%BeginExpansion
{\displaystyle\bigwedge\nolimits^{2}}
%EndExpansion
TM\hookrightarrow\mathcal{C\ell}_{(2)}^{(0)}(TM)\subset\mathcal{C\ell}%
^{(0)}(TM)$. In \cite{s1,s2,s3} the author identified erroneously $\mathbf{F}$
with an electromagnetic field. We discuss in detail that issue in a sequel
paper \cite{cliforms}. Now, we write the formula for $Q\otimes\tilde{Q}$ where
$Q\in\mathbb{C}(2)\otimes%
%TCIMACRO{\dbigwedge ^{1}}%
%BeginExpansion
{\displaystyle\bigwedge^{1}}
%EndExpansion
T^{\ast}M$ given by Eq.(\ref{S4}) is the matrix representation of
$\mathbf{Q}\in\sec\mathcal{C\ell}_{(0,2)}^{(0)}(M)\otimes%
%TCIMACRO{\dbigwedge ^{1}}%
%BeginExpansion
{\displaystyle\bigwedge^{1}}
%EndExpansion
T^{\ast}M$.

We have,%
\begin{align}
Q\otimes\check{Q}  &  =Q\overset{s}{\mathbf{\otimes}}\tilde{Q}+Q\wedge
\check{Q}\nonumber\\
&  =(\mathbf{-}g_{\mu\nu}dx^{\mu}\otimes dx^{v})\sigma_{0}+(\varepsilon
_{i\text{ }j}^{k}v_{\mu}^{i}(x)v_{\nu}^{j}(x)\text{ }dx^{\mu}\wedge
dx^{v})(-\mathrm{i}\sigma_{k})\nonumber\\
&  =-g\sigma_{0}+\mathbf{F}^{\prime k}\mathrm{i}\sigma_{k}, \label{33again}%
\end{align}
with
\begin{equation}
\mathbf{F}^{\prime k}=\frac{1}{2}\mathbf{F}_{\mu\nu}^{\prime k}dx^{\mu}\wedge
dx^{v}=\varepsilon_{i\text{ }j}^{k}v_{\mu}^{i}(x)v_{\nu}^{j}(x)dx^{\mu}\wedge
dx^{\nu}. \label{36}%
\end{equation}

For future reference we also introduce
\begin{equation}
\mathbf{F}_{\mu\nu}^{\prime}=\mathbf{F}_{\mu\nu}^{\prime k}\mathrm{i}%
\sigma_{k}. \label{37}%
\end{equation}

\subsection{Covariant Derivatives of Spinor Fields}

We now briefly recall the concept of covariant spinor derivatives
\cite{choquet,lami,moro,28}. The idea is the following:

(i) Every connection on the principal bundle of orthonormal frames
$P_{\mathrm{SO}_{1,3}^{e}}(M)$ determines in a canonical way a unique
connection on the principal bundle $P_{\mathrm{Spin}_{1,3}^{e}}(M)$.

(ii) Let $D$ be a covariant derivative operator acting on sections of an
associated vector bundle to $P_{\mathrm{SO}_{1,3}^{e}}(M)$, say, the tensor
bundle \ $\tau M$ and let $D^{s}$ be the corresponding covariant spinor
derivative \ acting on sections of associate vector bundles to
$P_{\mathrm{Spin}_{1,3}^{e}}(M)$, say, e.g., the spinor bundles $\mathcal{S}%
\left(  M\right)  $, $\mathcal{\dot{S}}(M)$ and $\mathcal{P}(M)\simeq
\mathcal{C\ell}^{(0)}(M)$, which may be called \textit{Pauli spinor bundle}.
The matrix representations of the above bundles are:%

\begin{align}
S\left(  M\right)   &  =P_{\mathrm{Spin}_{1,3}^{e}}(M)\times_{D^{(\frac{1}%
{2}0)}}\mathbb{C}^{2},\hspace{0.15in}\dot{S}(M)=P_{\mathrm{Spin}_{1,3}^{e}%
}(M)\times_{D^{(0,\frac{1}{2})}}\mathbb{C}_{2}\nonumber\\
P(M)  &  =S\left(  M\right)  \otimes\dot{S}(M)=P_{\mathrm{Spin}_{1,3}^{e}%
}(M)\times_{D^{(\frac{1}{2}0)}\otimes D^{(0,\frac{1}{2})}}\mathbb{C}%
^{2}\otimes\mathbb{C}_{2}, \label{D1}%
\end{align}
and $P(M)$ may be called \textit{matrix} \textit{Pauli spinor bundle}. Of
course, $P(M)\simeq\mathbb{C}\ell^{(0)}(M)$.

(iii) We have for $\mathbf{T}\in\sec%
%TCIMACRO{\dbigwedge }%
%BeginExpansion
{\displaystyle\bigwedge}
%EndExpansion
TM\hookrightarrow\mathcal{C\ell}^{(0)}(M)$ and $%
%TCIMACRO{\TeXButton{xi}{\mbox{\boldmath{$\xi$}}}}%
%BeginExpansion
\mbox{\boldmath{$\xi$}}%
%EndExpansion
\in\sec\mathcal{S}(M)$, $\overset{\cdot}{%
%TCIMACRO{\TeXButton{xi}{\mbox{\boldmath{$\xi$}}}}%
%BeginExpansion
\mbox{\boldmath{$\xi$}}%
%EndExpansion
}\in\sec\mathcal{\dot{S}}(M)$, $\mathbf{P}\in\sec$ $\mathcal{P}(M)$and
$\ \mathbf{v}\in\sec TM$ ,%

\begin{align}
D_{\mathbf{v}}^{s}(\mathbf{T}\otimes%
%TCIMACRO{\TeXButton{xi}{\mbox{\boldmath{$\xi$}}}}%
%BeginExpansion
\mbox{\boldmath{$\xi$}}%
%EndExpansion
)  &  =D_{\mathbf{v}}\mathbf{T}\otimes%
%TCIMACRO{\TeXButton{xi}{\mbox{\boldmath{$\xi$}}}}%
%BeginExpansion
\mbox{\boldmath{$\xi$}}%
%EndExpansion
+\mathbf{T\otimes}D_{\mathbf{v}}^{s}%
%TCIMACRO{\TeXButton{xi}{\mbox{\boldmath{$\xi$}}}}%
%BeginExpansion
\mbox{\boldmath{$\xi$}}%
%EndExpansion
,\nonumber\\
D_{\mathbf{v}}^{s}(\mathbf{T}\otimes\overset{\cdot}{%
%TCIMACRO{\TeXButton{xi}{\mbox{\boldmath{$\xi$}}}}%
%BeginExpansion
\mbox{\boldmath{$\xi$}}%
%EndExpansion
})  &  =D_{\mathbf{v}}\mathbf{T}\otimes\overset{\cdot}{%
%TCIMACRO{\TeXButton{xi}{\mbox{\boldmath{$\xi$}}}}%
%BeginExpansion
\mbox{\boldmath{$\xi$}}%
%EndExpansion
}+\mathbf{T\otimes}D_{\mathbf{v}}^{s}\overset{\cdot}{%
%TCIMACRO{\TeXButton{xi}{\mbox{\boldmath{$\xi$}}}}%
%BeginExpansion
\mbox{\boldmath{$\xi$}}%
%EndExpansion
}, \label{D.2}%
\end{align}
where (see \cite{28} for details)%
\begin{align}
D_{\mathbf{v}}\mathbf{T}  &  =\partial_{\mathbf{v}}\mathbf{T}+\frac{1}{2}[%
%TCIMACRO{\TeXButton{omega}{\mbox{\boldmath{$\omega$}}}}%
%BeginExpansion
\mbox{\boldmath{$\omega$}}%
%EndExpansion
_{\mathbf{v}},\mathbf{T}],\nonumber\\
D_{\mathbf{v}}^{s}%
%TCIMACRO{\TeXButton{xi}{\mbox{\boldmath{$\xi$}}}}%
%BeginExpansion
\mbox{\boldmath{$\xi$}}%
%EndExpansion
&  =\partial_{\mathbf{v}}%
%TCIMACRO{\TeXButton{xi}{\mbox{\boldmath{$\xi$}}}}%
%BeginExpansion
\mbox{\boldmath{$\xi$}}%
%EndExpansion
+\frac{1}{2}%
%TCIMACRO{\TeXButton{omega}{\mbox{\boldmath{$\omega$}}}}%
%BeginExpansion
\mbox{\boldmath{$\omega$}}%
%EndExpansion
_{\mathbf{v}}%
%TCIMACRO{\TeXButton{xi}{\mbox{\boldmath{$\xi$}}}}%
%BeginExpansion
\mbox{\boldmath{$\xi$}}%
%EndExpansion
,\hspace{0.15in}\nonumber\\
D_{\mathbf{v}}^{s}\overset{\cdot}{%
%TCIMACRO{\TeXButton{xi}{\mbox{\boldmath{$\xi$}}}}%
%BeginExpansion
\mbox{\boldmath{$\xi$}}%
%EndExpansion
}  &  =\partial_{\mathbf{v}}\overset{\cdot}{%
%TCIMACRO{\TeXButton{xi}{\mbox{\boldmath{$\xi$}}}}%
%BeginExpansion
\mbox{\boldmath{$\xi$}}%
%EndExpansion
}-\frac{1}{2}\overset{\cdot}{%
%TCIMACRO{\TeXButton{xi}{\mbox{\boldmath{$\xi$}}}}%
%BeginExpansion
\mbox{\boldmath{$\xi$}}%
%EndExpansion
}%
%TCIMACRO{\TeXButton{omega}{\mbox{\boldmath{$\omega$}}}}%
%BeginExpansion
\mbox{\boldmath{$\omega$}}%
%EndExpansion
_{\mathbf{v}},\nonumber\\
D_{\mathbf{v}}P  &  =\partial_{\mathbf{v}}\mathbf{P}+\frac{1}{2}%
%TCIMACRO{\TeXButton{omega}{\mbox{\boldmath{$\omega$}}}}%
%BeginExpansion
\mbox{\boldmath{$\omega$}}%
%EndExpansion
_{\mathbf{v}}\mathbf{P}-\frac{1}{2}\mathbf{P}\text{ }%
%TCIMACRO{\TeXButton{omega}{\mbox{\boldmath{$\omega$}}}}%
%BeginExpansion
\mbox{\boldmath{$\omega$}}%
%EndExpansion
_{\mathbf{v}}=\partial_{\mathbf{v}}\mathbf{P}+\frac{1}{2}[%
%TCIMACRO{\TeXButton{omega}{\mbox{\boldmath{$\omega$}}}}%
%BeginExpansion
\mbox{\boldmath{$\omega$}}%
%EndExpansion
_{\mathbf{v}},\mathbf{P}]. \label{D.4}%
\end{align}

(iv) For $\mathbf{T}\in\sec%
%TCIMACRO{\dbigwedge }%
%BeginExpansion
{\displaystyle\bigwedge}
%EndExpansion
TM\hookrightarrow\mathcal{C\ell}^{(0)}(TM)$ and $\xi\in\sec S(M)$, $\bar{\xi
}\in\sec\bar{S}(M)$, $P\in\sec$ $P(M)$and $\ \mathbf{v}\in\sec TM$ , we have
\begin{align}
D_{\mathbf{v}}^{s}(\mathbf{T}\otimes\xi)  &  =D_{\mathbf{v}}\mathbf{T}%
\otimes\xi+\mathbf{T}D_{\mathbf{v}}^{s}\xi,\label{D.2'}\\
D_{\mathbf{v}}^{s}(\mathbf{T}\otimes\bar{\xi})  &  =D_{\mathbf{v}}%
\mathbf{T}\otimes\bar{\xi}+\mathbf{T}D_{\mathbf{v}}^{s}\bar{\xi}\nonumber
\end{align}
\ \ 

and (see \cite{28} for details)%
\begin{align}
D_{\mathbf{v}}\mathbf{T}  &  =\partial_{\mathbf{v}}\mathbf{T}+\frac{1}{2}[%
%TCIMACRO{\TeXButton{omega}{\mbox{\boldmath{$\omega$}}}}%
%BeginExpansion
\mbox{\boldmath{$\omega$}}%
%EndExpansion
_{\mathbf{v}},\mathbf{T}],\nonumber\\
D_{\mathbf{v}}^{s}\xi &  =\partial_{\mathbf{v}}\xi+\frac{1}{2}\Omega
_{\mathbf{v}}\xi,\hspace{0.15in}\nonumber\\
D_{\mathbf{v}}^{s}\dot{\xi}  &  =\partial_{\mathbf{v}}\dot{\xi}-\frac{1}%
{2}\dot{\xi}\Omega_{\mathbf{v}},\nonumber\\
D_{\mathbf{v}}P  &  =\partial_{\mathbf{v}}P+\frac{1}{2}\Omega_{\mathbf{v}%
}P-\frac{1}{2}P\text{ }\Omega_{\mathbf{v}}=\partial_{\mathbf{v}}P+\frac{1}%
{2}[\Omega_{\mathbf{v}},P]. \label{D.4'}%
\end{align}

In the above equations $%
%TCIMACRO{\TeXButton{omega}{\mbox{\boldmath{$\omega$}}}}%
%BeginExpansion
\mbox{\boldmath{$\omega$}}%
%EndExpansion
_{\mathbf{v}}\in\sec\mathcal{C\ell}^{(0)}(TM)$ and $\Omega_{\mathbf{v}}\in
\sec$ $P(M)$. Writing as usual, $\mathbf{v}=v^{\mathbf{a}}\mathbf{e}%
_{\mathbf{a}}$, \ $D_{\mathbf{e}_{\mathbf{a}}}e^{\mathbf{b}}=-\mathbf{\omega
}_{\mathbf{ac}}^{\mathbf{b}}e^{\mathbf{c}}$ , $\mathbf{\omega}_{\mathbf{abc}%
}=-\mathbf{\omega}_{\mathbf{cba}}=\eta_{\mathbf{ad}}\mathbf{\omega
}_{\mathbf{bc}}^{\mathbf{d}},$ $\mathbf{\omega}_{\text{ \textbf{b}}%
}^{\mathbf{a}\text{ }\mathbf{c}}$ $=-\mathbf{\omega}_{\text{ \textbf{b}}%
}^{\mathbf{c}\text{ }\mathbf{a}}$, $\sigma_{\mathbf{b}}=\mathbf{e}%
_{\mathbf{b}}\mathbf{e}_{\mathbf{0}}$ and\footnote{Have in mind that $i$ is a
\textit{Clifford field} here.}\textsl{ i }$=-%
%TCIMACRO{\TeXButton{sigma}{\mbox{\boldmath{$\sigma$}}}}%
%BeginExpansion
\mbox{\boldmath{$\sigma$}}%
%EndExpansion
_{\mathbf{1}}%
%TCIMACRO{\TeXButton{sigma}{\mbox{\boldmath{$\sigma$}}}}%
%BeginExpansion
\mbox{\boldmath{$\sigma$}}%
%EndExpansion
_{\mathbf{2}}%
%TCIMACRO{\TeXButton{sigma}{\mbox{\boldmath{$\sigma$}}}}%
%BeginExpansion
\mbox{\boldmath{$\sigma$}}%
%EndExpansion
_{\mathbf{3}}$, we have
\begin{align}%
%TCIMACRO{\TeXButton{omega}{\mbox{\boldmath{$\omega$}}}}%
%BeginExpansion
\mbox{\boldmath{$\omega$}}%
%EndExpansion
_{\mathbf{e}_{\mathbf{a}}}  &  =\frac{1}{2}\mathbf{\omega}_{\mathbf{a}%
}^{\mathbf{bc}}\mathbf{e}_{\mathbf{b}}\mathbf{e}_{\mathbf{c}}=\frac{1}%
{2}\mathbf{\omega}_{\mathbf{a}}^{\mathbf{bc}}\mathbf{e}_{\mathbf{b}}%
\wedge\mathbf{e}_{\mathbf{c}}\nonumber\\
&  =\frac{1}{2}\mathbf{\omega}_{\mathbf{a}}^{\mathbf{bc}}%
%TCIMACRO{\TeXButton{sigma}{\mbox{\boldmath{$\sigma$}}}}%
%BeginExpansion
\mbox{\boldmath{$\sigma$}}%
%EndExpansion
_{\mathbf{b}}\overset{\vee}{%
%TCIMACRO{\TeXButton{sigma}{\mbox{\boldmath{$\sigma$}}}}%
%BeginExpansion
\mbox{\boldmath{$\sigma$}}%
%EndExpansion
}_{\mathbf{c}}\hspace{0.15in}\hspace{0.15in}\nonumber\\
&  =\frac{1}{2}(-2\mathbf{\omega}_{\mathbf{a}}^{\mathbf{0i}}%
%TCIMACRO{\TeXButton{sigma}{\mbox{\boldmath{$\sigma$}}}}%
%BeginExpansion
\mbox{\boldmath{$\sigma$}}%
%EndExpansion
_{\mathbf{i}}+\mathbf{\omega}_{\mathbf{a}}^{\mathbf{ji}}%
%TCIMACRO{\TeXButton{sigma}{\mbox{\boldmath{$\sigma$}}}}%
%BeginExpansion
\mbox{\boldmath{$\sigma$}}%
%EndExpansion
_{\mathbf{i}}%
%TCIMACRO{\TeXButton{sigma}{\mbox{\boldmath{$\sigma$}}}}%
%BeginExpansion
\mbox{\boldmath{$\sigma$}}%
%EndExpansion
_{\mathbf{j}})\nonumber\\
&  =\frac{1}{2}(-2\mathbf{\omega}_{\mathbf{a}}^{\mathbf{0i}}%
%TCIMACRO{\TeXButton{sigma}{\mbox{\boldmath{$\sigma$}}}}%
%BeginExpansion
\mbox{\boldmath{$\sigma$}}%
%EndExpansion
_{\mathbf{i}}-\text{\textsl{i}}\ \varepsilon_{\mathbf{i}\text{ }\mathbf{j}%
}^{\mathbf{k}}\mathbf{\omega}_{\mathbf{a}}^{\mathbf{ji}}%
%TCIMACRO{\TeXButton{sigma}{\mbox{\boldmath{$\sigma$}}}}%
%BeginExpansion
\mbox{\boldmath{$\sigma$}}%
%EndExpansion
_{\mathbf{k}})=%
%TCIMACRO{\TeXButton{Omega}{\mbox{\boldmath{$\Omega$}}}}%
%BeginExpansion
\mbox{\boldmath{$\Omega$}}%
%EndExpansion
_{\mathbf{a}}^{\mathbf{b}}\sigma_{\mathbf{b}}. \label{D.5}%
\end{align}

Note that the $%
%TCIMACRO{\TeXButton{Omega}{\mbox{\boldmath{$\Omega$}}}}%
%BeginExpansion
\mbox{\boldmath{$\Omega$}}%
%EndExpansion
_{\mathbf{a}}^{\mathbf{b}}$ are \ `formally' complex numbers. Also, observe
that we can write the \ `formal' Hermitian conjugate $%
%TCIMACRO{\TeXButton{omega}{\mbox{\boldmath{$\omega$}}}}%
%BeginExpansion
\mbox{\boldmath{$\omega$}}%
%EndExpansion
_{\mathbf{e}_{\mathbf{a}}}^{\dagger}$ of $%
%TCIMACRO{\TeXButton{omega}{\mbox{\boldmath{$\omega$}}}}%
%BeginExpansion
\mbox{\boldmath{$\omega$}}%
%EndExpansion
_{\mathbf{e}_{\mathbf{a}}}$ as
\begin{equation}%
%TCIMACRO{\TeXButton{omega}{\mbox{\boldmath{$\omega$}}}}%
%BeginExpansion
\mbox{\boldmath{$\omega$}}%
%EndExpansion
_{\mathbf{e}_{\mathbf{a}}}^{\dagger}=-\mathbf{e}^{\mathbf{0}}%
%TCIMACRO{\TeXButton{omega}{\mbox{\boldmath{$\omega$}}}}%
%BeginExpansion
\mbox{\boldmath{$\omega$}}%
%EndExpansion
_{\mathbf{e}_{\mathbf{a}}}\mathbf{e}^{\mathbf{0}}. \label{D.5'}%
\end{equation}
Also, write $\Omega_{\mathbf{e}_{\mathbf{a}}}$ for the matrix representation
of $%
%TCIMACRO{\TeXButton{omega}{\mbox{\boldmath{$\omega$}}}}%
%BeginExpansion
\mbox{\boldmath{$\omega$}}%
%EndExpansion
_{\mathbf{e}_{\mathbf{a}}}$, i.e.,
\[
\Omega_{\mathbf{e}_{\mathbf{a}}}=\Omega_{\mathbf{a}}^{\mathbf{b}}%
\sigma_{\mathbf{b}},
\]
where $\Omega_{\mathbf{a}}^{\mathbf{b}}$ are complex numbers with the same
coefficients as the \ `formally' complex numbers $\mathbf{\Omega}_{\mathbf{a}%
}^{\mathbf{b}}$. We can easily verify that
\begin{equation}
\Omega_{\mathbf{e}_{\mathbf{a}}}=\varepsilon\Omega_{\mathbf{e}_{\mathbf{a}}%
}^{\dagger}\varepsilon. \label{D.7}%
\end{equation}

We can prove the third line of Eq.(\ref{D.4'}) as follows. First, take the
Hermitian conjugation of \ the second line of Eq.(\ref{D.4'}), obtaining%
\[
D_{\mathbf{v}}\bar{\xi}=\partial_{\mathbf{v}}\bar{\xi}+\frac{1}{2}\bar{\xi
}\Omega_{\mathbf{v}}^{\dagger}.\hspace{0.15in}%
\]
Next multiply the above equation on the left by $\varepsilon$ and recall that
$\dot{\xi}=\bar{\xi}\varepsilon$ and Eq.(\ref{D.7}). We get
\begin{align*}
D_{\mathbf{v}}\dot{\xi}  &  =\partial_{\mathbf{v}}\dot{\xi}-\frac{1}{2}%
\dot{\xi}\varepsilon\Omega_{\mathbf{v}}^{\dagger}\varepsilon\\
&  =D_{\mathbf{v}}\dot{\xi}=\partial_{\mathbf{v}}\dot{\xi}-\frac{1}{2}\dot
{\xi}\Omega_{\mathbf{v}}.
\end{align*}
Note that this is compatible with the identification $\mathcal{C\ell}%
^{(0)}(TM)\simeq\mathcal{S}(M)\otimes_{\mathbb{C}}\mathcal{\dot{S}}(M)$ and
$\mathbb{C}\ell^{(0)}(M)\simeq S(M)\otimes_{\mathbb{C}}\dot{S}(M)$.

Note moreover that \ if $\mathbf{q}_{\mu}=e_{\mu}\mathbf{e}_{\mathbf{0}%
}=h_{\mu}^{\mathbf{a}}\mathbf{e}_{\mathbf{a}}\mathbf{e}_{\mathbf{0}}=h_{\mu
}^{\mathbf{a}}%
%TCIMACRO{\TeXButton{sigma}{\mbox{\boldmath{$\sigma$}}}}%
%BeginExpansion
\mbox{\boldmath{$\sigma$}}%
%EndExpansion
_{\mathbf{a}}$ $\in\mathcal{C\ell}^{(0)}(TM)\simeq\mathcal{S}(M)\otimes
_{\mathbb{C}}\mathcal{\dot{S}}(M)$ we have, \ %

\begin{equation}
D_{\mathbf{v}}\mathbf{q}_{\mu}=\partial_{\mathbf{v}}q_{\mu}+\frac{1}{2}%
%TCIMACRO{\TeXButton{omega}{\mbox{\boldmath{$\omega$}}}}%
%BeginExpansion
\mbox{\boldmath{$\omega$}}%
%EndExpansion
_{\mathbf{v}}\mathbf{q}_{\mu}+\frac{1}{2}\mathbf{q}_{\mu}%
%TCIMACRO{\TeXButton{omega}{\mbox{\boldmath{$\omega$}}}}%
%BeginExpansion
\mbox{\boldmath{$\omega$}}%
%EndExpansion
_{\mathbf{v}}^{\dagger}. \label{D.08}%
\end{equation}
For $q_{\mu}=h_{\mu}^{\mathbf{a}}\sigma_{\mathbf{a}}$ $\in\sec\mathbb{C}%
\ell^{(0)}(M)\simeq S(M)\otimes_{\mathbb{C}}\bar{S}(M),$ the matrix
representative of the $\mathbf{q}_{\mu}$ we have for any vector field
$\mathbf{v}\in\sec TM$ \
\begin{equation}
D_{\mathbf{v}}q_{\mu}=\partial_{\mathbf{v}}q_{\mu}+\frac{1}{2}\Omega
_{\mathbf{v}}q_{\mu}+\frac{1}{2}q_{\mu}\text{ }\Omega_{\mathbf{v}}^{\dagger}
\label{D.8}%
\end{equation}
which is the equation used by Sachs for the \textit{spinor} covariant
derivative of his \ `quaternion' fields. Note that M. Sachs in \cite{s1,s3}
introduced also a kind of total covariant derivative for his would be
`quaternion' fields. That \ `derivative' denoted in this text by
$D_{\mathbf{v}}^{\mathbf{S}}$ will be discussed below.

\subsection{Geometrical Meaning of $D_{e_{\nu}}q_{\mu}=\Gamma_{\nu\mu}%
^{\alpha}q_{\alpha}$}

We recall that Sachs wrote \footnote{See, e.g., Eq.(3.69) in \cite{s1}.}
without any mathematically justified argument that
\begin{equation}
D_{e_{\nu}}q_{\mu}=\Gamma_{\nu\mu}^{\alpha}q_{\alpha}, \label{D.9}%
\end{equation}
where $\Gamma_{\nu\mu}^{\alpha}$ are the connection coefficients of the
coordinate basis $\{e_{\mu}\}$, i.e.,
\begin{equation}
D_{e_{\nu}}e_{\mu}=\Gamma_{\nu\mu}^{\alpha}e_{\alpha}. \label{D.9bis}%
\end{equation}

How, can Eq.(\ref{D.9}) be true? Well, let us calculate $D_{e_{\nu}}%
\mathbf{q}_{\mu}$ in $\mathcal{C\ell}(TM)$. We have,%

\begin{align}
D_{e_{\nu}}\mathbf{q}_{\mu}  &  =D_{e_{\nu}}(e_{\mu}\mathbf{e}_{\mathbf{0}%
})\nonumber\\
&  =(D_{e_{\nu}}e_{\mu})\mathbf{e}_{\mathbf{0}}+e_{\mu}(D_{e_{\nu}}%
\mathbf{e}_{\mathbf{0}})\nonumber\\
&  =\Gamma_{\nu\mu}^{\alpha}\mathbf{q}_{\alpha}+e_{\mu}(D_{e_{\nu}}%
\mathbf{e}_{\mathbf{0}}). \label{D.10}%
\end{align}

So, Eq.(\ref{D.9}) follows if, and only if
\begin{equation}
D_{e_{\nu}}\mathbf{e}_{\mathbf{0}}=0. \label{D.11}%
\end{equation}

To understand the physical meaning of Eq.(\ref{D.11}) let us recall the
following. In Relativity Theory \textit{reference frames} are represented by
time like vector fields $Z\in\sec TM$ pointing to the future
\cite{rosharif,sw}. If we write the $\alpha_{\mathbf{Z}}=g(\mathbf{Z,)\in}%
%TCIMACRO{\dbigwedge \nolimits^{1}}%
%BeginExpansion
{\displaystyle\bigwedge\nolimits^{1}}
%EndExpansion
T^{\ast}M$ for the physically equivalent 1-form field, we have the well known
\emph{decomposition}
\begin{equation}
D\alpha_{\mathbf{Z}}=\mathbf{a}_{\mathbf{Z}}\otimes\alpha_{\mathbf{Z}%
}+\mathbf{\varpi}_{\mathbf{Z}}+\mathbf{\sigma}_{\mathbf{Z}}+\frac{1}%
{3}E_{\mathbf{Z}}\mathbf{p}, \label{D.12}%
\end{equation}
where
\begin{equation}
\mathbf{p}=g-\alpha_{\mathbf{Z}}\otimes\alpha_{\mathbf{Z}} \label{D.13}%
\end{equation}
is called the projection tensor (and gives the metric of the rest space of an
instantaneous observer \cite{sw}), $\mathbf{a}_{\mathbf{Z}}=g(D_{\mathbf{Z}%
}\mathbf{Z,)}$ is the (form) acceleration of $\mathbf{Z}$, $\mathbf{\varpi
}_{\mathbf{Z}}$ is the rotation of $\mathbf{Z}$, $\mathbf{\sigma}_{\mathbf{Z}%
}$ is the shear of $\mathbf{Z}$ and $E_{\mathbf{Z}}$ is the expansion ratio of
$\mathbf{Z}$ . In a coordinate chart ($U,x^{\mu}$), writing $\mathbf{Z}%
=Z^{\mu}\partial/\partial x^{\mu}$ \ and $\mathbf{p}=(g_{\mu\nu}-Z_{\mu}%
Z_{\nu})dx^{\mu}\otimes dx^{\nu}$ we have
\begin{align}
\mathbf{\varpi}_{\mathbf{Z}\mu\nu}  &  =Z_{\left[  \alpha;\beta\right]
}p_{\mu}^{\alpha}p_{\nu}^{\beta},\nonumber\\
\mathbf{\sigma}_{\mathbf{Z}\alpha\beta}  &  =[Z_{\left(  \mu;\nu\right)
}-\frac{1}{3}\mathbf{E}_{\mathbf{Z}}h_{\mu\nu}]p_{\alpha}^{\mu}p_{\beta}^{\nu
},\nonumber\\
E_{\mathbf{Z}}  &  =Z^{\mu};_{\mu}. \label{D.14}%
\end{align}

Now, in Special Relativity where the space time manifold is the structure
$(M\mathcal{=}\mathbb{R}^{4},$ $g=\eta,D^{\eta},\tau_{\eta},\uparrow
)$\footnote{$\mathbf{\eta}$ is a constant metric, i.e., there exists a chart
$\langle x^{\mu}\rangle$ of $\ M=\mathbb{R}^{4}$ such that $\mathbf{\eta
}(\partial/\partial x^{\mu},\partial/\partial x^{\nu})=\eta_{\mu\nu}$, the
numbers $\eta_{\mu\nu}$ forming a diagonal matrix with entries $(1,-1,-1,-1)$.
Also, $D^{\mathbf{\eta}}$ is the Levi-Civita connection of $\mathbf{\eta}$.}
an \emph{inertial reference frame }(\emph{IRF}) $\mathbf{I}\in\sec TM$ is
defined by $D^{\eta}\mathbf{I}=0$. We can show very easily (see, e.g.,
\cite{sw}) that in General Relativity Theory $\left(  \emph{GRT}\right)  $
\ where each gravitational field is modelled by a spacetime\footnote{More
precisely, by a diffeomorphism equivalence class of Lorentzian spacetimes,
according to current dogma.} $(M,g,D,\tau_{g},\uparrow)$ there is \textit{in
general} no shear free frame $(\sigma_{\mathfrak{Q}}=0)$ on any open
neighborhood $U$ \ of any given spacetime point. The reason is clear if we use
local coordinates $\langle x^{\mu}\rangle$ covering $U$. Indeed,
$\sigma_{\mathfrak{Q}}=0$ implies five independent conditions on the
components of the frame $\mathfrak{Q}$. Then, we arrive at the conclusion that
in a general spacetime model\footnote{We take the opportunity to correct an
statement in \cite{rosharif}. There it is stated that in General Relativity
there are no inertial frames. Of, course, the correct statement is that in a
general spacetime model there are \textit{in general} no inertial frames. But,
of course, there are spacetime models where there exist frames $\mathfrak{Q}%
\in$ $\sec TU\subset\sec TM$ satisfying \ $D\mathfrak{Q}=0$. See below.} there
is no frame $\mathfrak{Q}\in$ $\sec TU\subset\sec TM$ satisfying
\ $D\mathfrak{Q}=0$, and in general there is no \emph{IRF} in any model of
\emph{GRT}. Saying that, if there exists in a model of General Relativity a
frame $\mathfrak{Q}$ satisfying $D\mathfrak{Q}=0$, we agree in calling
$\mathfrak{Q}$ \ an inertial frame.

The following question arises naturally: which characteristics a reference
frame on a \emph{GRT} spacetime model must have in order to reflect as much as
possible the properties of an \emph{IRF} of \emph{SRT}?

The answer to that question \cite{rosharif} is that there are two kind of
frames in \emph{GRT} such that each frame in one of these classes share some
important aspects of the \emph{IRFs} of \emph{SRT}. Both concepts are useful
and it is important to distinguish between them in order to avoid
misunderstandings. These frames are the \emph{pseudo inertial} \emph{reference
frame }(\emph{PIRF}) and the and the local Lorentz reference frames
(\emph{LLRF}$\gamma$\emph{s}), but we don not need to enter the details here.

On the open set $U\subset M$ covered by a coordinate chart $\langle x^{\mu
}\rangle$ of the maximal atlas of $M$ multiplying Eq.(\ref{D.11}) by
$h_{\mathbf{a}}^{\nu}$ \ such that $\mathbf{e}_{\mathbf{a}}=h_{\mathbf{a}%
}^{\nu}e_{\nu}$, we get%
\begin{equation}
D_{\mathbf{e}_{\mathbf{a}}}\mathbf{e}_{\mathbf{0}}=0;\hspace{0.15cm}%
\mathbf{a}=0\mathbf{,}1,2,3.\text{ } \label{D.psi}%
\end{equation}

Then, it follows that
\begin{equation}
D_{X}\mathbf{e}_{\mathbf{0}}=0\text{, }\forall X\in\sec TM \label{D.new}%
\end{equation}
which characterizes $\mathbf{e}_{\mathbf{0}}$ as an inertial frame. This
imposes several restrictions on the spacetime described by the theory. Indeed,
if Eq.(\ref{D.new}) holds, we must have
\begin{equation}
\mathbf{Ric}(\mathbf{e}_{\mathbf{0}},X)=0\text{, }\forall X\in\sec TM,
\label{D.neww}%
\end{equation}
where, $\mathbf{Ric}$ is the Ricci tensor of the manifold modelling spacetime
\footnote{See, exercise 3.2.12 of \cite{sw}.}. In particular, this condition
cannot be realized in Einstein-de Sitter spacetime. This fact is completely
hidden in the matrix formalism used in M. Sachs theory, where no restriction
on the spacetime manifold (besides the one of being a spin manifold) need to
be imposed.

\subsection{Geometrical Meaning of $D_{e_{\mu}}\sigma_{\mathbf{i}}=0$ in
General Relativity}

We now discuss what happens in the usual theory of dotted and undotted two
component \textit{matrix} spinor fields in general relativity, as \ described,
e.g., in \cite{carmeli,penrose,penrindler}. In that formulation it is
\textit{postulated} that the covariant spinor derivative of Pauli matrices
must satisfy
\begin{equation}
D_{e_{\mu}}\sigma_{\mathbf{i}}=0,\hspace{0.15cm}\mathbf{i=}1,2,3. \label{D.15}%
\end{equation}

Eq.(\ref{D.15}) translate in our formalism as%
\begin{equation}
D_{e_{\mu}}%
%TCIMACRO{\TeXButton{sigma}{\mbox{\boldmath{$\sigma$}}}}%
%BeginExpansion
\mbox{\boldmath{$\sigma$}}%
%EndExpansion
_{\mathbf{i}}=D_{e_{\mu}}\left(  \mathbf{e}_{\mathbf{i}}\mathbf{e}%
_{\mathbf{0}}\right)  =0. \label{D.16}%
\end{equation}
Differently from the case of Sachs theory, Eq.(\ref{D.16}) can be satisfied
if
\begin{equation}
D_{e_{\mu}}\mathbf{e}_{\mathbf{i}}=\mathbf{e}_{\mathbf{i}}(D_{e_{\mu}%
}\mathbf{e}_{0})\mathbf{e}_{\mathbf{0}} \label{D.15bis}%
\end{equation}
or, writing $D_{e_{\mu}}\mathbf{e}_{\mathbf{a}}=\omega_{\mu\mathbf{a}%
}^{\mathbf{b}}\mathbf{e}_{\mathbf{b}}$, we have
\begin{equation}
\omega_{\mu\mathbf{i}}^{\mathbf{b}}=\mathbf{e}^{\mathbf{b}}\lrcorner
(\omega_{\mu\mathbf{0}}^{\mathbf{a}}\mathbf{e}_{\mathbf{i}}\mathbf{e}%
_{\mathbf{a}}\mathbf{e}_{\mathbf{0}}), \label{D.15biss}%
\end{equation}
where $\lrcorner$ is the left contraction operator in the Clifford bundle
(see, e.g., \cite{28}, for details). This certainly implies some restrictions
on possible spacetime models, but that is the price, necessary to be paid, in
order to have spinor fields. At least we do not need to necessarily have
$D\mathbf{e}_{\mathbf{0}}=0$.

We analyze some possibilities of satisfying Eq.(\ref{D.15}):

(\textbf{i}) Suppose that $\mathbf{e}_{\mathbf{0}}$ satisfy $D_{e_{\mu}%
}\mathbf{e}_{\mathbf{0}}=0$, i.e., $D\mathbf{e}_{\mathbf{0}}=0$.\textit{
}Then, a necessary and sufficient condition for the validity of Eq.(\ref{D.16}%
) is that%
\begin{equation}
D_{e_{\mu}}\mathbf{e}_{\mathbf{i}}=0. \label{D.17}%
\end{equation}

\qquad Multiplying Eq.(\ref{D.17}) by $h_{\mathbf{a}}^{\mu}$ we get \
\begin{equation}
D_{\mathbf{e}_{\mathbf{a}}}\mathbf{e}_{\mathbf{i}}=0,\mathbf{\hspace
{0.15cm}i=}1,2,3;\hspace{0.15cm}\mathbf{a}=0,1,2,3. \label{D.newbis}%
\end{equation}

In particular,
\begin{equation}
D_{\mathbf{e}_{0}}\mathbf{e}_{\mathbf{i}}=0,\hspace{0.15cm}\mathbf{i=}1,2,3.
\label{D.18}%
\end{equation}

Eq.(\ref{D.18}) means that the fields $\mathbf{e}_{i}$ following each integral
line $\sigma$ of $\mathbf{e}_{\mathbf{0}}$ are Fermi transported\footnote{An
original approach to the Fermi transport using Clifford bundle methods has
been given in \cite{rovapa}. There an equivalent spinor equation to the famous
Darboux equations of differential geometry is derived.} \cite{sw}. Physicists
interpret that equation saying that the $\left.  \mathbf{e}_{i}\right\vert
_{\sigma(I)}$ are physically realizable by gyroscopic axes, which gives the
local standard of \textit{no} rotation.

The above conclusion sounds fine. However it follows from Eq.(\ref{D.new}) and
Eq.(\ref{D.newbis}) that
\begin{equation}
D_{\mathbf{e}_{\mathbf{a}}}\mathbf{e}_{\mathbf{b}}=0,\hspace{0.15cm}%
\mathbf{a=}0\mathbf{,}1,2,3;\hspace{0.15cm}\mathbf{b}=0,1,2,3. \label{D.19}%
\end{equation}

Recalling that existence of spinor fields implies that $\{\mathbf{e}%
_{\mathbf{a}}\}$ is a global tetrad \cite{g1}, Eq.(\ref{D.19}) implies that
the connection $D$ must be teleparallel. Then, under the above conditions the
curvature tensor of a spacetime admitting spinor fields must be\textit{ null}.
This, is in particular, the case of Minkowski spacetime.

(\textbf{ii}) Suppose now that $\mathbf{e}_{\mathbf{0}}$ is a geodesic frame,
i.e., $D_{\mathbf{e}_{0}}\mathbf{e}_{\mathbf{0}}=0$. Then, $h_{\mathbf{0}%
}^{\nu}D_{e_{\nu}}\mathbf{e}_{0}=0$ and Eq. (\ref{D.15bis}) implies only that
\
\begin{equation}
D_{\mathbf{e}_{0}}\mathbf{e}_{i}=0;\hspace{0.15cm}i\mathbf{=}1,2,3
\label{D.18bis}%
\end{equation}
If we take an integral line of $\mathbf{e}_{\mathbf{0}}$, say $\gamma$, then
the set $\{\left.  \mathbf{e}_{\mathbf{a}}\right\vert _{\gamma}\}$ may be
called an \textit{inertial moving frame} along $\gamma$. The set $\{\left.
\mathbf{e}_{\mathbf{a}}\right\vert _{\gamma}\}$ is also \emph{Fermi}
transported (as can be easily verified) since $\gamma$ is a geodesic
worldline. They define the standard of \emph{no} rotation along $\gamma.$

In conclusion, a consistent definition of spinor fields in General Relativity
using the Clifford and spin-Clifford bundles formalism of this paper needs not
only the triviality of the frame bundle, i.e., existence of a global tetrad,
say $\{\mathbf{e}_{\mathbf{a}}\}$. It also needs the validity of
Eq.(\ref{D.15bis}). A nice physical interpretation follows moreover if the
tetrad satisfies
\begin{equation}
D_{\mathbf{e}_{0}}\mathbf{e}_{\mathbf{a}}=0;\hspace{0.15cm}\mathbf{a=}%
0\mathbf{,}1,2,3. \label{D.18biss}%
\end{equation}

Of course, as it is the case in Sachs theory, the matrix formulation of spinor
fields do not impose any constrains in the possible spacetime models, besides
the one needed for the existence of a spinor structure. Saying that we have an
important comment, presented in the next section.

\subsection{Covariant Derivative of the Dirac Gamma Matrices}

If we use a real spin bundle where we can formulate the Dirac equation, e.g.,
one where the typical fiber is the ideal of (algebraic) Dirac spinors, i.e.,
the ideal generated by a idempotent $\frac{1}{2}(1+E_{0})$, $E_{0}%
\in\mathbb{R}_{1,3}$, $E_{0}\cdot E_{0}=1,$ then no restriction is imposed on
the global tetrad field $\{\mathbf{e}_{\mathbf{a}}\}$ defining the spinor
structure of spacetime (see \cite{28,moro}). In particular, since%
\begin{equation}
D_{\mathbf{e}_{a}}\mathbf{e}_{\mathbf{b}}=\omega_{\mathbf{ab}}^{\mathbf{c}%
}\mathbf{e}_{\mathbf{c}}, \label{Dirac1}%
\end{equation}
we have,
\begin{equation}
D_{\mathbf{e}_{a}}\mathbf{e}_{\mathbf{b}}=\frac{1}{2}[%
%TCIMACRO{\TeXButton{omega}{\mbox{\boldmath{$\omega$}}}}%
%BeginExpansion
\mbox{\boldmath{$\omega$}}%
%EndExpansion
_{\mathbf{e}_{\mathbf{a}}},\mathbf{e}_{\mathbf{b}}] \label{Dirac2}%
\end{equation}

Then,
\begin{equation}
\omega_{\mathbf{ab}}^{\mathbf{c}}\mathbf{e}_{\mathbf{c}}-\frac{1}{2}%
%TCIMACRO{\TeXButton{omega}{\mbox{\boldmath{$\omega$}}}}%
%BeginExpansion
\mbox{\boldmath{$\omega$}}%
%EndExpansion
_{\mathbf{e}_{\mathbf{a}}}\mathbf{e}_{\mathbf{b}}+\frac{1}{2}\mathbf{e}%
_{\mathbf{b}}%
%TCIMACRO{\TeXButton{omega}{\mbox{\boldmath{$\omega$}}}}%
%BeginExpansion
\mbox{\boldmath{$\omega$}}%
%EndExpansion
_{\mathbf{e}_{\mathbf{a}}}=0. \label{Dirac3}%
\end{equation}

The matrix representation of the real spinor bundle, of course, sends
$\{\mathbf{e}_{\mathbf{a}}\}\mapsto\{\gamma_{\mathbf{a}}\}$, where the
$\gamma_{\mathbf{a}}$'s are the standard representation of the Dirac matrices.
Then, the matrix translation of Eq.(\ref{Dirac3}) is
\begin{equation}
\omega_{\mathbf{ab}}^{\mathbf{c}}\gamma_{\mathbf{c}}-\frac{1}{2}%
\omega_{\mathbf{e}_{\mathbf{a}}}\gamma_{\mathbf{b}}+\frac{1}{2}\gamma
_{\mathbf{b}}\omega_{\mathbf{e}_{\mathbf{a}}}=0. \label{Dirac4}%
\end{equation}

For the matrix elements $\gamma_{\mathbf{b}B}^{A}$ we have
\begin{equation}
\omega_{\mathbf{ab}}^{\mathbf{c}}\gamma_{\mathbf{c}B}^{A}-\frac{1}{2}%
\omega_{\mathbf{e}_{\mathbf{a}}C}^{A}\gamma_{\mathbf{b}B}^{C}+\frac{1}%
{2}\gamma_{\mathbf{b}C}^{A}\omega_{\mathbf{e}_{\mathbf{a}}B}^{C}=0.
\label{Dirac5}%
\end{equation}

In \cite{choquet} this last equation is confused with the covariant derivative
of $\gamma_{\mathbf{c}B}^{A}$. Indeed in an exercise in problem 4, Chapter
Vbis \cite{choquet} ask one to prove that

\begin{center}%
\begin{tabular}
[c]{|l|}\hline
$\nabla_{\mathbf{e}_{\mathbf{b}}}\gamma_{\mathbf{c}B}^{A}=\omega_{\mathbf{ab}%
}^{\mathbf{c}}\gamma_{\mathbf{c}B}^{A}-\frac{1}{2}\omega_{\mathbf{e}%
_{\mathbf{a}}C}^{A}\gamma_{\mathbf{b}B}^{C}+\frac{1}{2}\gamma_{\mathbf{b}%
C}^{A}\omega_{\mathbf{e}_{\mathbf{a}}B}^{C}=0.$\\\hline
\end{tabular}

\end{center}

Of course, the first member of the above equation does not define any
covariant derivative operator. Confusions as that one appears over and over
again in the literature, and of course, is also present in Sachs theory in a
small modified form, as shown in the next subsubsection.

\subsection{$D_{e_{\nu}}^{\mathbf{S}}q_{\mu}=0$}

Taking into account Eq.(\ref{D.8}) and Eq.(\ref{D.9}) we can write:%
\begin{equation}
\partial_{\nu}\mathbf{q}_{\mu}+\frac{1}{2}%
%TCIMACRO{\TeXButton{omega}{\mbox{\boldmath{$\omega$}}}}%
%BeginExpansion
\mbox{\boldmath{$\omega$}}%
%EndExpansion
_{\mathbf{\nu}}\mathbf{q}_{\mu}+\frac{1}{2}\mathbf{q}_{\mu}%
%TCIMACRO{\TeXButton{omega}{\mbox{\boldmath{$\omega$}}}}%
%BeginExpansion
\mbox{\boldmath{$\omega$}}%
%EndExpansion
_{\mathbf{\nu}}-\Gamma_{\nu\mu}^{\alpha}\mathbf{q}_{\alpha}=0. \label{D.18'}%
\end{equation}

Write,
\begin{equation}
D_{e_{\nu}}^{\mathbf{S}}\mathbf{q}_{\mu}=\partial_{\nu}\mathbf{q}_{\mu}%
+\frac{1}{2}%
%TCIMACRO{\TeXButton{omega}{\mbox{\boldmath{$\omega$}}}}%
%BeginExpansion
\mbox{\boldmath{$\omega$}}%
%EndExpansion
_{\mathbf{\nu}}\mathbf{q}_{\mu}+\frac{1}{2}\mathbf{q}_{\mu}%
%TCIMACRO{\TeXButton{omega}{\mbox{\boldmath{$\omega$}}}}%
%BeginExpansion
\mbox{\boldmath{$\omega$}}%
%EndExpansion
_{\mathbf{\nu}}-\Gamma_{\nu\mu}^{\alpha}\mathbf{q}_{\alpha} \label{D.18''}%
\end{equation}
from where
\begin{equation}
D_{e_{\nu}}^{\mathbf{S}}\mathbf{q}_{\mu}=0. \label{D.18'''}%
\end{equation}
Of course, the matrix representation of the last two equations are:%
\begin{align}
D_{e_{\nu}}^{\mathbf{S}}q_{\mu}  &  =\partial_{\nu}q_{\mu}+\frac{1}{2}%
\Omega_{\mathbf{\nu}}q_{\mu}+\frac{1}{2}q_{\mu}\text{ }\Omega_{\mathbf{\nu}%
}^{\dagger}-\Gamma_{\nu\mu}^{\alpha}q_{\alpha}.\nonumber\\
D_{e_{\nu}}^{\mathbf{S}}q_{\mu}  &  =0. \label{D.19'}%
\end{align}

Sachs call \footnote{See Eq.(3.69) in \cite{s1}.} $D_{e_{\nu}}^{\mathbf{S}%
}q_{\mu}$ the covariant derivative of a $q_{\mu}$ field. The nomination is an
\textit{unfortunate} one, since the equation $D_{e_{\nu}}^{\mathbf{S}}q_{\mu
}=0$ is a \textit{ trivial} identity and do not introduce any new connection
in the game.\footnote{The equation $D_{\nu}^{\mathbf{S}}\mathbf{q}_{\mu}=0$
(or its matrix representation) is a reminicescence of an analogous equation
for the components of tetrad fields often printed in physics textbooks and
confused with the metric compatibility condition of the connection. See,e.g.,
comments on page 76 of \cite{goschu}.}

After this long exercise we can derive easily all formulas in chapters 3-6 of
\ \cite{s1} without using \textit{any} matrix representation at all. In
particular, for use in the sequel paper \cite{cliforms} we collect some formulas,%

\begin{align}
\mathbf{q}^{\mu}\mathbf{\check{q}}_{\mu}  &  =-4,\hspace{0.15in}q^{\mu}%
\check{q}_{\mu}=-4\sigma_{0}\nonumber\\
\mathbf{q}_{\rho}^{\mu}%
%TCIMACRO{\TeXButton{omega}{\mbox{\boldmath{$\omega$}}}}%
%BeginExpansion
\mbox{\boldmath{$\omega$}}%
%EndExpansion
\mathbf{\check{q}}_{\mu}  &  =0,\hspace{0.15in}q^{\mu}\Omega_{\rho}\check
{q}_{\mu}=0,\nonumber\\%
%TCIMACRO{\TeXButton{omega}{\mbox{\boldmath{$\omega$}}}}%
%BeginExpansion
\mbox{\boldmath{$\omega$}}%
%EndExpansion
_{\rho}  &  =-\frac{1}{2}\mathbf{\check{q}}_{\mu}(\partial_{\rho}%
\mathbf{q}^{\mu}+\Gamma_{\rho\tau}^{\mu}\mathbf{q}^{\tau}),\hspace
{0.15in}\Omega_{\rho}=-\frac{1}{2}\check{q}_{\mu}(\partial_{\rho}q^{\mu
}+\Gamma_{\rho\tau}^{\mu}q^{\tau}) \label{D.20}%
\end{align}

\qquad As a last remark, please keep in mind that our \ `normalization' of $%
%TCIMACRO{\TeXButton{omega}{\mbox{\boldmath{$\omega$}}}}%
%BeginExpansion
\mbox{\boldmath{$\omega$}}%
%EndExpansion
_{\rho}$ (and of $\Omega_{\rho}$) here \textit{differs} from Sachs one by a
factor of $1/2$. We prefer our normalization, since it is more natural and
avoid factors of $2$ when we perform contractions.

\section{Conclusions}

In this paper we \ recalled the concept of covariant derivatives of algebraic
dotted and undotted spinor fields, when these objects are represented as
sections of real spinor bundles (\cite{lami,moro,28}) and study how this
theory has as matrix representative the standard spinor fields (dotted and
undotted) already introduced long ago, see, e.g., \cite{carmeli, penrose,
penrindler, pirani}. Through our approach is that was possible to identify a
profound physical meaning concerning some of the rules used in the standard
formulation of the (matrix) formulation of spinor fields, e.g., why the
covariant derivative of the Pauli matrices must be null. Those rules implies
in \textit{constraints} for the geometry of the spacetime manifold. A possible
realization of that constraints is one where the fields defining a global
tetrad must be such that $\mathbf{e}_{\mathbf{0}}$ is a geodesic field and the
$\left.  \mathbf{e}_{i}\right\vert $ $_{\gamma\text{ }}$are Fermi transported
(i.e., are not rotating relative to the local gyroscopes axes) along each
integral line $\gamma$ of $\mathbf{e}_{\mathbf{0}}$. For the best of our
knowledge this important fact is here disclosed for the first time.

We use our formalism to disclose the mathematical \textit{nature }of the basic
variables of Sachs "unified" theory as discussed recently in \cite{s3} and as
originally introduced in \cite{s1}. \ More on that theory will be discussed in
a sequel paper \cite{cliforms}.

\begin{acknowledgement}
Authors are grateful to Dr. Ricardo A. Mosna for very useful observations.
\end{acknowledgement}

\end{document}